
\input jnl.tex
\input tableofc.tex
\input reforder.tex
\input eqnorder.tex
\input tables.tex
\input defs.tex

\def\slanta{\raise.15ex\hbox{$/$}\kern-.57em\hbox{$a$}}

\def\orodo{\raise-.01ex\hbox{O}\kern-.85em\hbox{$\prod$}}

\def\orodo{\raise-.01ex\hbox{O}\kern-.85em\hbox{$\prod$}}

\rightline{NSF-ITP-XX-XX}

version 8/26

\title Quantum Hall Fluids\footnote{$^*$}{To appear %
in the Proceedings of the South African School of Physics,
 Tsitsikamma 1994,  Springer-Verlag}

\author A. Zee

\affil Institute for Theoretical Physics
\& Department of Physics
University of California
Santa Barbara, California 93106-4030

\abstract{We review the effective field theory treatment of topological
quantum fluids, focussing on the Hall fluids.}

\endtopmatter

\head{ I. Topological Quantum Fluids }
\taghead{1.}

In the last few years, an interesting new subject, the study of
topological quantum fluids, has emerged. Examples of topological quantum
fluids include the Hall fluid, the chiral spin fluid, and the anyon superfluid.
I have lectured on the chiral
spin fluid and anyon superfluid in more details elsewhere\refto{losalamos,
braz},
to which I refer the interested reader. Here I will focus on the Hall fluid.
Much of the work described here were done in collaboration with X. G.
Wen. I will discuss here the effective field theory approach to the Hall
fluid. The lectures by X.G. Wen in this volume are focussed on the
microscopic theory of the Hall fluid and complement these lectures.
Some closely related topics are discussed by P. Wiegmann in his lectures in
this volume.

Without further ado we will now talk about the quantum Hall
system.\refto{GP}
 It is a
remarkably simple system to describe, just a bunch of electrons moving in
a plane in the
presence
of an external magnetic field $B$ perpendicular to  the plane. The
magnetic
field is assumed to be sufficiently strong so that the  spin of the electrons
may be completely polarized. We take all the electrons to  have spin up,
say, so
that the electrons are treated as spinless fermions.  As is well known by
now,
this seemingly innocuous and simple physical  situation contains a
fascinating
wealth of physics.

This remarkable richness
 follows from the interplay between two basic pieces of physics, as follows.

(1) Even though the electron is point-like, it takes up a finite amount
of room.

Classically, a charged particle  in a magnetic field moves in a
Larmor circle of radius $r$ determined by
$$
evB=mv^2/r
\eqno(fma)
$$
As you can see, I am really starting these
lectures with the basics: ``$F=ma$"! Classically, the  radius is not fixed.
The more energetic particles move in larger circles. Applying
semi-classical quantization
$$
mvr \sim h = 2\pi
\eqno(quant)
$$
(in units in which $\hbar$  is  equal to unity) we obtain
$$
mvr = eBr^2 \sim 2\pi
\eqno(larmor)
$$
Thus, the area taken up by each electron is of order
$$
\pi r^2 \sim 2\pi^2/eB
\eqno(area)
$$

(2) Electrons are fermions and want to stay out of each other's way.

According to Pauli, not only do electrons each insists on taking up a finite
amount of room, they have to have their own rooms. Thus, the quantum
Hall problem may be described as a sort of housing crisis, or as the
problem
of assigning office space at  the Institute for
Theoretical Physics to visitors who do not want to share offices.

The rich physics of the quantum Hall system is a consequence of these
two fundamental points. Already at this stage, we would expect that when
the
number of electrons $N_e$ is just right to fill out space completely, namely
when
$$
N_e \pi r^2  \sim  N_e(2\pi^2/eB) \sim A,
\eqno(just)
$$
we have a special situation.  Here $A$ denotes the area of the system.
The fact that this is somehow special should be
reflected in the physics.

These heuristic considerations should, and could, be made precise, of
course. First, let us solve the Schr\"odinger equation satisfied by a
single electron in a magnetic field
$$
-\Big[(\partial_x-ieA_x)^2+(\partial_y-ieA_y)^2\Big] \psi=2mE\psi
\eqno(schro)
$$
This is a textbook problem solved by Landau years ago. (See also Wen's
lectures in this volume.) The eigenvalues are given by
$$
E_n=\left(n+\12\right) eB/m,\qquad n=0,1,2,\ldots.\eqno(7.3)
$$
The (degenerate) set of states with the energy $E_n$ is referred to as the
$n$-th Landau level. The  levels are separated by the Landau-Larmor
energy
$$
E_L = \hbar\omega_c \equiv eB/m
\eqno(energy)
$$
where
$\omega_c$ is just the classical cyclotron frequency.

Each Landau level
has degeneracy $BA/2\pi$ where $A$ is the area of the
system. The
physical origin of this vast degeneracy is clear from the
preceding
discussion.   Because of
translation invariance the Larmor circles may be
placed anywhere. The
degeneracy is just the area of the system divided by
the area of the Larmor
circle. It is thus natural to define a filling factor
defined as the ratio of the
number of electrons to the
number of states in
a Landau level
$$\nu ={N_e\over (eBA/2\pi)} ={2\pi n_e\over
eB}\eqno(7.4)$$
where $n_e$ is the density of electrons.  Note that

$$\nu^{-1} ={(BA/N_e)\over 2\pi}\eqno(7.5)$$
is equal to the flux per
particle measured in fundamental units of flux.
(Here and in what follows,
we will often choose units such that $\hbar, c$,
and $e$ are set equal to 1.
In these units the fundamental
unit of flux is $hc/e=2\pi$.)

(To make the
definition of the filling factor precise, we have to discuss the
quantum
Hall system on a sphere rather than on a plane. On a sphere, the
number
of states in a Landau level is a discrete finite number. We will not
worry
about such details here. See section XIII for a discussion.)

Imagine filling the system with non-interacting
electrons. By the Pauli
exclusion principle, each succeeding electron we
put in has to go into a
different state in the first Landau level. Clearly
then, the situation when the
first Landau level is just filled is special. (This
corresponds to the situation
in which the Larmor circles just fills up the plane
in our heuristic argument.)
The next electron would have to go into the second Landau level.
To put in just one more electron would cost us
much more  energy than what we spent for the preceding electron.
The quantum Hall fluid with filling fraction
$\nu$ equal to 1, and more
generally equal to an integer, is
incompressible. Any attempt  to compress
a quantum Hall fluid lessens the
degeneracy of the Landau levels
 (the effective area $A$ decreases and so
the degeneracy ${BA\over2\pi}$
 decreases) and thus
some of the
electrons would have to be pushed to the next level, costing
 energy. This is the essence of the so-called integer
quantum
Hall effect.

Thus, the integer quantum Hall effect is relatively easy to
understand and
indeed was anticipated in the theoretical literature.\refto{ando}

(The
actual theory is
more
complicated and has yet to be worked out in all details. One has to
understand
how the simple picture given here is
modified by the presence of
impurities and by
the interaction between electrons. An understanding of
how impurities
localize the electron states is essential to understanding the
experimental
data, a subject that would take us far beyond the scope of these lectures.
Let us content ourselves with a brief caricature here. Impurities may be
treated as a bunch of short ranged potentials with random strength and
placed randomly. Some of the single electron wave functions are localized:
the electron is trapped in a potential well and its wave function falls off
exponentially. Some other wave functions are extended: they manage to
avoid the deeper potential wells and to wind their way around and between
the potential wells. Clearly, only electrons in the extended wave functions
can carry electric current from one side of the sample to the other and thus
contribute to the conductance. Thus, we expect a given Landau level to
broaden out into a band of nearly, but not quite, degenerate states. It turns
out that most of the states in the Landau band are localized except for some
extended states
in the middle. Thus, as we fill the system with electrons, the conductance
stays contant and then suddenly jumps as the fermi energy exceeds the
energy of the extended states in the middle of the band.)

In contrast, the experimental
discovery of the fractional Hall effect, namely
that the Hall fluid is also
incompressible for filling factor $\nu$ equal to
simple odd denominator
fractions such as 1/3 and 1/5, took theorists
completely by surprise. In our
simple picture, for filling factor equal to
1/3 for example, only one third
of the states in the first Landau level is
filled. It would seem that throwing
in a few more electrons would not have
that much effect on the system.
Why should the $\nu=1/3$ Hall fluid be also
incompressible? Here the interaction between electrons is crucial. Let us
fill the first Landau level with non-interacting spinless electron up to
$\nu=1/3$. The important point is that this does not define a unique
many-body state: there is an enormous degeneracy. Each of the electrons
can go
into any of the $BA/2\pi$ states available subject only to the constraint of
Pauli exclusion. But as soon as we turn on a repulsive interaction, a
presumably unique ground state is picked out, within the space of this vast
set of  degenerate states.

We have an important clue in trying to
 understand why Hall fluids with inverse filling factor $\nu^{-1}$ equal to
an odd integer are special. Notice that  $\nu =
N_e/N_{\phi}$.
For the fluids in question, the number of
flux quantum per electron is an odd integer. What does that tell us?

\head{ II. Quantum Phase and Fractional Statistics}
\taghead{2.}

Clearly, the interplay between the quantum of charge and the quantum of
flux is going to be
of central importance to our discussion.
 I assume you know about the celebrated
Aharonov-Bohm effect.\refto{ab} When a charged particle of charge $q$
goes
around a flux $\Phi$, the quantum wave function of the system acquires a
phase factor equal to
$$
e^{iq\Phi}
\eqno(ab)
$$
This effect forms the basis of the phenomenon of fractional
statistics.\refto{LM,FS,WZZ, ASWZ} By now it is well known how
fractional
statistics
can be implemented in a field theory context.\refto{WZZ} For later use in
these
lectures, let me remind you how this is done and at the same time I will end
up explaining to you what fractional statistics is.

Consider a theory in (2+1) dimensional spacetime with a conserved
current $j^\mu$. Let us couple
a gauge potential $a_\mu$ to this current and describe the dynamics of the
gauge potential by a so-called Chern-Simons term\refto{CSTX, CSTY, cs}
$$
\cL=2\gamma \eps^{\munu\la} a_\mu \part_\nu a_\la +a_\mu
j^\mu\eqno(4.6)$$
Here $\eps^{\munu\la}$ denotes the totally antisymmetric symbol in
(2+1) dimensional spacetime and $\gamma$ an arbitrary real parameter.
Note that under a gauge transformation $a_\mu \rightarrow a_\mu +
\part_\mu \La$, the Chern-Simons term changes by $\eps^{\munu\la}
a_\mu \part_\nu a_\la  \rightarrow \eps^{\munu\la} a_\mu \part_\nu
a_\la +
\eps^{\munu\la} \part_\mu \La \part_\nu a_\la$. The action changes by
$$
\delta S=2\gamma\int d^3x \eps^{\mu\nu\lambda}
\partial_\mu(\Lambda\partial_\nu
a_\lambda) \eqno(bou)
$$
Thus, if we are allowed to drop boundary terms, which we will assume to
be
the case for now,  the  Chern-Simons action is gauge invariant.
Later we will worry about situations in which we are not allowed to
drop the boundary terms.

Consider the equation of motion from \(4.6) then. We have
$$
4\gamma\eps^{\mu\nu\la}\partial_\nu a_\la =-j^\mu
\eqno(mot)
$$
Around a particle sitting at rest, the current $j_i$ vanishes. From \(mot)
we
can calculate the flux on the particle to be
$$
\int d^2x (\partial_1 a_2-\partial_2 a_1)=-{1\over4\gamma}\int d^2xj^0
\eqno(fx)
$$
Thus, the Chern-Simons term has the effect of endowing the charged
particles in the theory with flux. According to Aharonov-Bohm then, when
one of our particles moves around another, the wave function acquires a
phase. (Here charged particles simply means the
particles which couple to the gauge potential $a_\mu$. In this context,
when
we refer to charge and flux, we are not referring to the charge and flux
associated with the ordinary electromagnetic field. We are simply
borrowing a useful set of terminology.)

The existence of bosons and fermions represents one of the most profound
features of quantum physics. When we interchange two identical quantum
particles, the wave function acquires a factor of either $+1$ or $-1$. Note
that to
interchange two particles, we can move one of them half-way around the
other and then translate both of them by an appropriate distance. Thus,
when we interchange two particles described by \(4.6), the wave
function acquires an additional phase, in addition to whatever phase the
wave function would have acquired due to the ``intrinsic" quantum statistics
of the particles (fermi or bose), that is, whatever phase the wave function
would have acquired in the absence of the Chern-Simons term. The
particles are said to carry fractional statistics.

It is customary to characterize fractional statistics by a parameter $\theta$
such that when two particles both with statistics described by $\theta$ are
interchanged the wave function acquires a phase factor $e^{i\theta}$. I
leave
it to the student to work out that  for the
particles described in \(fx) the statistics parameter is given
by
$$\th={1\over 8\gamma}\eqno(4.7)$$
There is a somewhat tricky factor of two discussed in \Ref{FOT}. So if
you disagree with \(4.7) by a factor of two, don't despair. Try again.

Strictly speaking, the term ``fractional statistics" is
somewhat misleading. First, a trivial remark is that the statistics parameter
$\theta/\pi$ does not have to be a fraction. Here, statistics is not directly
related to counting.
The statistics between our particles is perhaps better thought of as a
long-ranged phase interaction between them, mediated by a gauge boson. In
any case, the effective statistics of particles living in (2+1) dimensional
spacetime may be other than bose or fermi. Particles carrying fractional
statistics are called anyons. Particles with $\theta=\pi/2$ are called semions.

An alternative treatment is to integrate out the gauge potential $a$
altogether. We can either functionally integrate out $a$ in the path integral
formalism or simply solve for $a$ in terms of $j$ and insert the result
in \(fx). Let me remind you how this goes. Given the Lagrangian
$$
\cL = \phi \cQ \phi + \phi J,
\eqno(toy)
$$
where $\cQ$ denotes some operator,
we have the equation of motion
$$
2 \cQ \phi = - J
\eqno(toy2)
$$
with the solution
$$
\phi = {-1\over 2\cQ} J
\eqno(toy3)
$$
Eliminating $\phi$ in the Lagrangian, we obtain immediately the effective
Lagrangian
$$
\cL = - J {1\over 4\cQ} J
\eqno(toy4)
$$

Applying this formalism to our Lagrangian \(4.6) we obtain the non-local
Lagrangian
$$
\cL = {-1\over 8\gamma} j ({1\over \epsilon \part} ) j
\eqno(hopf)
$$
This is known as the Hopf term and was described in \Ref{WZZ}. There is
a slight technical subtlety here: due to gauge invariance the inverse of the
differential operator $\epsilon \part$ is not defined. In other words,
it has a zero mode of the form ($\part_\lambda$ acting on anything
smooth),
that is,
$$
(\epsilon^{\mu\nu\lambda}
\partial_\nu ) \part_\lambda {\rm acting\ on\ anything\ smooth\/} = 0
\eqno(zmode)
$$
This just the usual complication associated with gauge theories: we
just fix the gauge. Choose the Lorentz gauge
$$
\part a = 0
\eqno(lorentz)
$$
Then the Hopf term becomes
$$
{\cL}\;=\; {1\over 8\gamma} j_\mu  \bigl( {\epsilon^{\mu\nu\lambda}
\partial_\nu \over  \partial^2 }\bigr) j_\lambda\,. \eqno(11.19)
$$

The fractional statistics parameter can be determined as follows. Consider
two particles, one sitting at rest, the other moving half-way around the one
sitting at rest. The current $j$ is then equal to the sum of two terms
describing the two particles respectively. Plugging into \(11.19) we obtain
the result for the fractional statistics parameter that we cite above in
\(4.7).

Field theorists have  greatly enthused over the Chern-Simons
term\refto{CSTX,CSTY, cs} in recent
years. Let us just mention several of its endearing properties that are
particularly relevant for our purposes.

 (1) For the Chern-Simons term, gauge invariance implies
Lorentz
invariance.  Condensed matter physics is of course non-relativistic. Thus,
{\it a
priori}, one may think that there are two separate Chern-Simons terms,
$\eps_{ij}a_i\part_0a_j$ and $\eps_{ij}a_0\part_ia_j$. Gauge invariance
under
$a_\mu\to a_\mu-\part_\mu\La$, however, implies that the two terms
must combine
into the Chern-Simons term $\eps^{\mu\nu\la} a_\mu\part_\nu a_\la$.  In
contrast, the Maxwell term would in general be non-relativistic, consisting
of
two terms, $(f_{0i})^2$ and $(f_{ij})^2$, with an arbitrary relative
coefficient between them
 (with $f_{\mu\nu}=\partial_\mu a_\nu-\partial_\nu a_\mu$ as usual).
This is one reason that relativistic field theorists find this subject of
topological Hall and anyon fluids so appealing. As long as we are interested
only in long distance physics we can ignore the Maxwell term (see the next
remark) and play with a relativistic theory.

 (2) The Chern-Simons term dominates the Maxwell term $f^2$ at
long
distances:  it contains only one derivative while the Maxwell term contains
two. Note that this picks out (2+1)-dimensional spacetime as special.  In
(3+1)-dimensional spacetime the generalization of the Chern-Simons term
is
$\eps\part a\part a$, the $f\tilde f$ term much studied in connection with
quantum chromodynamics,
it has the same scale dimension as the Maxwell term $f^2$.
In higher dimensions, the Chern-Simons term, for example $\eps a\part
a\part a$
in (4+1)-dimensional space, is less important at long distances than the
Maxwell term $f^2$.  (However, there is
 an alternative generalization of the Chern-Simons term to higher
dimensions.\refto{HG})

 (3) Another important property of the Chern-Simons term is its
topological character. We will postpone a discussion of this crucial point.

We also mention for later use that the Chern-Simons term gives a mass to
the gauge boson.\refto{CSTY} We all know that a gauge potential with
dynamics
described solely by the Maxwell term is massless. Consider the Lagrangian
with an added Chern-Simons term (written schematically)
$$
\cL \sim k a \epsilon \part a + {1\over g^2} f^2
\eqno(csm)
$$
The important point is that, as remarked above, the Chern-Simons term
contains one derivative, while the Maxwell term contains two. Thus, in
momentum space, the inverse propagator of the gauge boson of momentum
$q$ has the schematic form $ k q + q^2/ g^2$ where we suppressed all
Lorentz indices etc. The propagator has a pole at $q \sim g^2 k$ and
the gauge boson has a mass of order $g^2 k$,
proportional to the coefficient of the Chern-Simons term. I leave it as an
exercise for the reader to put in all the Lorentz indices and to show that the
apparent pole at $q = 0$ describes a gauge degree of freedom.

Incidentally, this clears up an apparent paradox. Some physicists puzzled by
fractional statistics have reasoned that since the gauge boson described by
the
Chern-Simons term is massive and hence short ranged, it can't possible
generate
fractional statistics, which is manifestly an infinite ranged interaction. No
matter how far the two particles we are interchanging are, the wave
function still acquires a phase. The resolution is that the information is in
fact propagated over infinite range by the $q = 0$ pole noted above. This
apparent paradox is intimately connected with the puzzlement many
physicists (including Bohr, for example) felt when they first heard of the
Aharonov-Bohm effect. How can a particle in a region with no magnetic
field whatsoever and arbitrarily far from the magnetic flux know about the
existence of the magnetic flux?

\head {III. Fractional Hall Fluids}
\taghead{3.}

With this preliminary about fractional statistics out of the way, we are now
ready to see why Hall fluids with inverse filling factor equal to an odd
integer might be special. Recall that in this case
the background magnetic field contains
an odd number of magnetic flux quanta per electron.

The first argument we cite is due to Jain.\refto{J1,J2} Imagine an adiabatic
process in which we somehow move some of the flux quanta so that $p$
units of flux are attached to each electron. For $p$ even,
the additional Dirac-Aharonov-Bohm phase associated with moving one
electron
around another is $e^{i \pi p} = 1$ and so the statistics of the electron is
unchanged. The electrons are now moving in a reduced magnetic field
$B_{\rm{eff}} = B - 2 \pi p n$ where $n$ is the number density of
electrons.
(Recall that in
 our convention,  the unit of flux is $2\pi$.)
The filling factor has been increased to $\nu_{\rm{eff}}$, given by
$\nu_{\rm{eff}}^{-1} = (B - 2 \pi p n) / 2 \pi n = \nu^{-1} - p$.
For $\nu_{\rm{eff}} = m$ an integer, we have $\nu^{-1} = p + m^{-1}$
and
$$
\nu \enskip = \enskip \frac{m}{m p + 1}\,. \eqno(11.9)
$$
\noindent Thus, fractional Hall systems with
$\nu = m/(m p +1)$ \ ($p$ even) may be adiabatically changed into an
integer Hall system with filling factor $m$, as was also
 emphasized by
Greiter and
Wilczek.\refto{GW} Note that this argument gives us more
than we had
hoped for. The case  we wanted to understand,
with $\nu^{-
1}=$ an  odd integer, is obtained for $m = 1$.

In this rather neat way, the
fractional Hall fluid, more precisely, a Hall
fluid with $\nu = m/(m p +1)$
with $m$ an integer and $p$ an even
integer, is related to the integer
Hall
fluid. We may thus want to argue that since the integer Hall fluid is
incompressible the fractional Hall fluid is also incompressible.

Another argument, historically earlier than Jain's
argument,  is due to Zhang, Hansson, and
Kivelson\refto{ZHG}. Imagine attaching all of the flux to the electrons.
Thus, each electron gets attached to it an odd number of flux quanta. By
our preceding argument, the electrons with the attached flux quanta
become bosons. We now have bosons moving in the absence of a
background
magnetic field.
Since bosons can condense accoding to Bose and Einstein,
we conclude that Hall fluids with inverse filling factor equal to an odd
integer is energetically favored.

At this point, I may interject a historical remark which may be of
interest to some readers. In early 1974, after I wrote the paper with
Wilczek on how fractional statistics may be implemented in field theory by
introducing the Chern-Simons term,\refto{WZZ}
 I gave talks at various institutions. I
had a transparency mentioning that since statistics could be transformed at
will in (2+1) dimensions, we can always represent a system of electrons
on the plane as bosons. A Nobel laureate in the audience was quite taken by
this remark; however, he  and I couldn't imagine how it could be any
possible
use. Several years had to pass before Zhang \etal\refto{ZHG} figured out
the
relevance of this remark for the Hall system. At another institution,
another Nobel laureate advised me after my talk to abandon this line of
work. He felt that this sort of ``mathematical fooling around" wouldn't do
me much good.

\head {IV. A General Approach To Hall Fluids}
\taghead{4.}

Now that we have determined the physics responsible for making Hall
fluids
with special filling factors incompressible, we are now ready to construct
the effective field theory describing Hall fluids.\refto{gir, WZR, bw, wniu,
effective, ezawa, FKT}
Before I do this, I will first
give you a slick argument indicating what the
effective field theory must be. I will take the ``high road" and proceed
completely from general principles.\refto{FZ}

Let us start by listing what we know about the Hall system.

(1) We are living in (2+1) dimensional spacetime (because the electrons
are restricted to a plane.)

(2) The electromagnetic current $J_\mu$ is conserved: $\part^\mu
J_\mu=0$.

These two statements are certainly indisputable, but when combined they
already tell us that
the current can be
written  as the curl
of a vector potential, \ie
$$J^\mu={1\over 2\pi}\eps^{\munu\la} \part_\nu
\a_\la \eqno(12.1)$$
The factor of $1/(2\pi)$ defines the normalization of
$\a_\mu$. We learned in school that in three dimensional spacetime, if the
divergence of something is zero, then that something is the curl of
something else. That is precisely what \(12.1) says. The only sophistication
here is that what we learned in school works in Minkowskian space as well
as Euclidean space --- that is just a matter of a few signs here and there.

We now note that when we transform  $\a_\mu$ by
$\a_\mu\to
\a_\mu-\part_\mu\La$,  the current is unchanged.
In other words, $\a_\mu$ is a gauge potential.

We did not go looking for a gauge potential, the gauge potential came
looking for us! There is no place to hide. The  existence of a
gauge potential follows from completely general
considerations.

(3) We want to describe the system field theoretically by an effective local
Lagrangian.

(4) We are only interested in the physics at long distances and large time,
that is, at small wave number and low frequency.

Indeed, a field theoretic description of a physical system
 may
be regarded as a means of organizing various aspects of the
relevant
physics in a systematic way according to their relative importance
at
long
distances and according to symmetries.
We classify terms in a field
theoretic Lagrangian according to
powers of derivatives, powers of the
field variables, and so
forth.  A general scheme for classifying terms
is
according to their (naive) scaling dimensions.  In
(2+1) dimensional
spacetime, the current has mass dimension 2, and thus
the
gauge potential
$\a_\mu$ has dimension 1, as is the case for any gauge
potential coupled to
matter fields according to the gauge principle.

(5) Parity and time reversal are broken by the external magnetic field.

This last statement is just as indisputable as the statements in (1) and (2).
The experimentalist produces
the magnetic field by driving a current through a coil with the current
flowing
either clockwise or
anti-clockwise.

I will now show that given these five general statements we can deduce the
form of the effective Lagrangian.

Since gauge
invariance
forbids the dimension 2 term $\a_\mu \a^\mu$ in the
Lagrangian,
the
simplest possible term is in fact the dimension 3 Chern-Simons
term
$\eps^{\munu\la}\a_\mu\part_\nu\a_\la$. Thus, the Lagrangian
(density) is
simply
$$\cL={k\over 4\pi}\alpha \epsilon \part \alpha+\ldots\eqno(12.2)$$
where $k$ is a dimensionless
coefficient that we have yet to determine.

We have
introduced
and will use henceforth the compact notation
$\epsilon \alpha \partial \beta \equiv \epsilon^{\mu \nu \lambda}\,
\alpha_\mu \partial_\nu \beta_\lambda = \epsilon \beta \partial \alpha$
for two vector fields
$\alpha_\mu$ and $\beta_\mu$, where $\epsilon^{\mu\nu\lambda}$
denotes
the
antisymmetric symbol in (2+1) dimensional spacetime.

The
 terms indicated
by
($\ldots$) in \(12.2) include the dimension 4 Maxwell term ${1\over
g^2}(f_{0i}^2 - f_{ij}^2)$
and other terms with
higher
dimensions.  The important observation
is
that these higher dimensional terms are
less important at long distances.
The long distance physics is determined
purely by the Chern-Simons term,
(provided $k\not=0$ of course).

Note that in
general the coefficient $k$ may well be zero,
in which
case the
physics
is determined by the short distance
terms represented by the
$(\ldots)$ in \(12.2).
Put differently, a Hall  fluid may be defined as a two
dimensional electron
system for which the coefficient of the Chern-Simons
term does not vanish,
and
consequently is such that its long distance
physics is largely independent of
the microscopic details that define the
system.
Indeed, we may classify two dimensional electron systems
according to
whether
$k$ is zero or not.

To determine $k$ we
couple the system to an ``external'' or ``additional''
electromagnetic gauge
potential $A_\mu$.    From \(12.1) and \(12.2) we
now have
$$\cL={k\over 4\pi}
\eps^{\mu\nu\la} \a_\mu \part_\nu \a_\la - {1\over
2\pi}
\eps^{\munu\la} A_\mu
\part_\nu \a_\la={k\over 4\pi}
\eps^{\munu\la} \a_\mu \part_\nu \a_\la
-{1\over 2\pi} \eps^{\munu\la}
\a_\mu\part_\nu A_\la \eqno(12.4)$$
(In writing the second equality sign,
we have dropped a total derivative.)
Note that the gauge potential of the
magnetic field responsible for the Hall
effect should not be included in
$A_\mu$; it is implicitly contained already
in
the coefficient $k$.  We
can now
integrate out $\a$ to obtain an effective Lagrangian
for $A$.
We simply plug in \(hopf)
$$ {\pi\over k}
 j_\mu
\eps^{\munu\la} \part_\nu {1\over\part^2} j_\la \eqno(12.3)$$
and recognize that in \(12.3)
the current $J_\mu$ is just ${1\over 2\pi} \eps^{\munu\la}
\part_\nu
A_\la$ from \(12.4).
We obtain $$\cL_{\rm eff}
=-
{1\over 4\pi k} \eps^{\munu\la} A_\mu\part_\nu A_\la.
\eqno(12.5)$$
The electromagnetic current that flows in response to
$A_\mu$ is thus
$$J^\mu\equiv -{\de\cL_{\rm eff}
\over\de A_\mu}
={1\over 2\pi k} \eps^{\munu\la}
\part_\nu A_\la .\eqno(12.6)$$
The
time component of this equation tells us that an excess
density $\de n$
of
electrons is related to a local fluctuation of the magnetic field by
$$\de
n={1\over 2\pi k} \de B.\eqno(12.7)$$
Since we are describing a system
with filling factor $\nu=1$, we should
have
$k$ equal to 1.

We will see later that there is a mathematical argument showing that $k$
must be an integer.

Thus far,
we have discussed only non-interacting electrons.  Let us now
introduce
interactions between the electrons.  We have just argued that
the Chern-Simons term is the only possible term at long distances.  Thus,
the
only
effect of the interaction can be to modify the coefficient of the
Chern-Simons term.

We find it convenient to immediately
 study a more
general situation. Consider
non-interacting electrons at a density such that
they fill $m$ Landau levels:
$\nu=2\pi n/B=m$.
We argue that the current
of the electrons belonging to each
Landau level is separately conserved.
Physically, the large gap between the
Landau levels implies that it is a
good approximation to
take  the separate levels as dynamically
independent.
Thus, we introduce $m$ gauge potentials  as in
\(12.1)
$$J_I={1\over 2\pi} \eps\part\a_I\eqno(12.8)$$
for
$I=1,2,\ldots,m$.
We have suppressed indices.
  The Lagrangian \(12.2) is
then generalized to
$$4\pi \cL=\sum_I \a_I
\eps\part\a_I\eqno(12.9)$$

We now take into account the interaction
between the electrons.  The
interaction couples the different Landau levels,
or equivalently the $\a_I$'s.
We insist on the
physical argument that the
electrons ultimately can only interact
electromagnetically.  The interaction
should only involve the total
electromagnetic current
$J=\sum J_I=
{1\over 2\pi} \eps\part \left(\sum \a_I\right)$,
not the individual currents
$J_I$.
Thus, interaction can only change the Lagrangian to
$$4\pi
\cL=\sum_I \a_I \eps\part \a_I+p (\sum_I \a_I) \eps\part
(\sum_J
\eps\part\a_J),\eqno(12.10)$$
where $p$ is some real constant to be
determined.
Or more compactly, we can write the Lagrangian as
$$
{\cL}\;=\; \frac{1}{4\pi}\; \alpha K \varepsilon \partial \alpha
\;\eqno(11.17)
$$
\noindent Here $\alpha$ now denotes the column or row vector
$(\alpha_1, \alpha_2, \dots, \alpha_m)$ and we introduce the $m$ by $m$
matrix
$$
K = \left(\matrix{
p+1    & p      & \ldots & p     \\
p      & p+1    &        & .     \\
\vdots &        & \ddots & \vdots \\
p      & \ldots & \ldots & p+1\cr}
\right)
\eqno(11.16)
$$
or more concisely
$$K \equiv I + p \;C\eqno(kmatrix)$$ where $C$ is the matrix in which
every entry is
equal to 1. The real number $k$ in \(12.2) has now been promoted to a
matrix $K$.

The notion of quasiparticles or ``elementary" excitations is basic to
condensed matter physics. The effects of  many-body interaction may be
such that the quasi-particles in the system are no longer  the electrons
that make up the system at the microscopic level. Here we define the
quasiparticles as the entities which couple to the gauge potential. Thus,
we add to
the Lagrangian in \(11.16)
$$
\alpha j = \sum_I \alpha_I^\mu j_{I\mu}
\eqno(qua)
$$
As we will see later, these quasiparticles may be identified as vortices in
the Hall fluid. We will sometimes referred to them as vortices.

Just as  in \(12.4) we may couple the system to an external electromagnetic
potential $A$, so that
  after integration by parts we add to the Lagrangian the term
$$
-\sum_I {1\over 2\pi} \eps^{\munu\la}
\a_{I\mu}\part_\nu A_\la
\eqno(electro)
$$
Thus, we can incorporate these two couplings rather compactly by
introducing an effective current $\tilde{j}_I$
as the $I$-th quasiparticle or vortex current corrected by the
external electromagnetic field:
$$
\tilde{j}_{I\mu}\;=\; j_{I\mu}^{\rm{vortex}}\;-\;
\frac{1}{2\pi} \, \varepsilon_{\mu\nu \lambda}\, \partial^\nu A^\lambda
\,. \eqno(11.18)
$$
Thus, finally we obtain
$$
\cL={1\over4\pi}\alpha K\varepsilon \partial \alpha +\alpha j
\eqno(cop)
$$
Integrating out the gauge fields entirely we obtain a matrix
version of the Hopf interaction\refto{WZZ}
$$
{\cL}\;=\; \pi \tilde{j}_\mu K^{-1} \bigl( {\epsilon^{\mu\nu\lambda}
\partial_\nu \over \partial^2}\bigr) \tilde{j}_\lambda\,. \eqno(compact)
$$

We now have ``the world's most compact"  representation  of the
fractional quantum Hall fluid. Here $K$ denotes  an $m$-by-$m$ matrix
in
which an unknown  number $p$ appears. Everything else is fixed. At this
point, it may also be helpful to remind the reader that for $m=1$ this
effective Lagrangian simplifies drastically: there is only one gauge
potential and the matrix $K$ collapses into a number $(p+1)^{-1}$.

We can now easily read off the physics from \(compact).
The Lagrangian
 contains three types of terms: $AA, Aj$, and $jj$.
The $AA$ term has the schematic form $A {\epsilon \part \epsilon \part
\epsilon
\part \over \part^2} A$. Using $\epsilon \part  \epsilon \part \sim \part^2$
and cancelling between numerator and denominator, we obtain
$$\cL=-{1\over 4\pi} \eps A\part A \left(\sum_{IJ} K_{IJ}^{-1}\right).
\eqno(11.20)$$ Since $A$ does not know about the index $I$ the
dependence on the matrix $K$ can only factorize into the form shown.
Compare this with \(12.5). Just as before, we vary
with respect to $A$ to find the electromagnetic current
$$J_{em}^\mu ={1\over 2\pi} \eps^{\munu\la} \part_\nu A_\la
\left(\sum_{IJ}
K_{IJ}^{-1}\right).\eqno(11.21)$$
The $\mu=i$ components of this equation tell us that an
 electric field  produces a current in the orthogonal direction.  The
Hall conductance is thus
$$\si_H=\sum_{IJ} K_{IJ}^{-1}.\eqno(11.22)$$
equal to the filling factor $\nu$.

The $Aj$ term has the schematic form $j {\epsilon \part \epsilon \part
\over \part^2} A$. Cancelling the differential operators, we find
$$\cL=A^\mu\sum_{IJ} K_{IJ}^{-1} j_{J_\mu}.\eqno(11.23)$$
Thus, the electric charge of the $I$th type of vortices is determined to be
$$q_I=\sum_J K_{JI}^{-1}.\eqno(11.24)$$

Finally, the vortices interact with each other via
$$\cL=\pi\sum_{IJ} j_I^\mu K_{IJ}^{-1} {\eps_{\munu\la}
\part_\nu \over \part^2} j_J.
\eqno(11.25)$$  We simply remove the twiddle sign in \(11.19a). Referring
to our discussion in section II we see that a
bound state of vortices consisting of
$l_I$ vortices of the $I$th type would
have statistics given by
$${\th\over\pi} =\sum_{IJ}  l_I K_{IJ}^{-1} l_J.\eqno(11.26)$$

The three formulas \(11.22), \(11.24), and \(11.26)
 for the Hall conductance, the vortex charge,  and vortex
statistics, hold for any $K$. Before we try to invert  $K$ let us
apply these formulas to the simplest case of $m=1$, in which, as the reader
may recall, the matrix $K$ collapses to a $1$-by-$1$ matrix equal to
$(p+1)$.
Matrix inversion is now trivial. We find instantly  that
$$\eqalignno{\si_H&={1\over p+1},&(11.28')\cr
 q&={1\over p+1},&(11.29')\cr
\noalign{\hbox{  and}}
 {\th\over\pi}& ={1\over p+1}.&(11.30')\cr}$$

By now, the reader may well be wondering that while all this is fine and
good, what would actually allow us to determine this unknown constant
$p$? We would
now like to argue that the electron
or hole should appear
somewhere in
the excitation spectrum.\refto{TW}
After all, the theory is supposed
to
describe a system of electrons and thus far our rather general
Lagrangian
does
not contain
any reference to the electron!

Let us now
look for the hole (or electron).
We note from \(11.29')
that a
bound object made up of $(p+1)$ vortices would have charge equal
to 1. This is
perhaps the hole. For this to work, we see that $p$ has to be an integer.

What is the statistics of this bound state? Let us move one of these bound
objects half way around another such bound object, thus effectively
interchanging them. When each vortex moves around another we pick up a
phase given by  ${\th\over\pi} ={1\over (p+1)}$ according to
\(11.30'). But here we have $(p+1)$ vortices going around $(p+1)$
vortices and so we pick up a phase
$${\th\over\pi} ={1\over (p+1)}
(p+1)^2=(p+1).\eqno(12.11)$$
For the hole to be a fermion we must require
${\th\over\pi}$ to be an odd
integer.  This fixes $p$ to be an even integer.

Referring to \(11.28') we see that we are describing here fractional Hall
fluids with filling factor $\nu=1/(p+1)$ with $p=$ an even integer,
precisely the classic Laughlin odd-denominator Hall fluids. From \(11.29')
and \(11.30') we recover the famous result that the quasi-particles carry
fractional charge and statistics.

It is actually not difficult to invert the matrix
$K$ we have in \(11.16). Using the fact that $C$ is proportional to a
projection
$(C^2=mC)$, we can readily invert $K=I+pC$ to find
$$K^{-1} = I-\left({p\over 1+mp}\right) C.\eqno(11.27)$$

We find immediately that
$$\eqalignno{ \si_H &= {m\over mp+1},&(11.28)\cr
q_I& ={1\over mp+1},&(11.29)\cr
\noalign{\hbox{and}}
{\th\over\pi} &= \left(1-{p\over mp+1}\right) \left(\sum_I l_I\right)^2
&(11.30)\cr}$$
Note that $q_I$ is independent of $I$ (this would not be true for the more
general $K$ matrices we will discuss below.)
We will leave it to the reader
to work out that in this more complicated case also $p$ must be an even
integer.
(A general treatment will be
given later.)

Let us end this section with a couple of comments. Suppose we refuse to
introduce gauge potentials. Since the current $J_\mu$ has dimension 2,
the simplest term
constructed out of the currents, $J_\mu
J^\mu$,
 is already of dimension 4; indeed,
this is just the Maxwell term.
There is no way of constructing a dimension
3
local interaction out of the
currents directly.  To lower the dimension we are forced to introduce the
inverse of the derivative and write schematically $J {1\over \epsilon \part }
J$, which is of course just the non-local Hopf term. Thus, the question
``why gauge field?"
people often asked may be
answered in part by saying that the introduction of gauge
fields allows us to avoid
dealing with non-local interactions.

What we have given in this section is certainly a very slick derivation of
the effective long distance theory of the Hall fluid. Some would say too
slick. Let us go back to our five general statements or principles. Of these
five, four are absolutely indisputable. In fact, the most questionable is the
statement that looks the most innocuous to the casual reader, namely (3). In
general, the effective Lagrangian for a condensed matter system would be
non-local. We are implicitly assuming that the system does not
contain massless field the exchange of which would lead to non-local
interaction. Also implicit in (3) is the assumption that the Lagrangian can
be expressed completely in terms of the gauge potential $\alpha$. A priori,
we certainly do not know that there might not be other relevant degrees of
freedom. The point is that as long as these degrees of freedom are not
gapless they can be integrated out.

\head {V. Detailed Construction of Field Theory of Hall Fluids}
\taghead{5.}

For those who find the derivation given above
too slick, we
will now give a much ``sweatier" derivation of the field theory of Hall
fluids, making a series of approximations justified at each step by saying
that we are interested only in the long distance physics. We will implement
the physical picture described earlier in connection with Jain's argument.
Let us start
with the Lagrangian describing spinless electrons in a magnetic
field in the second quantized formalism:
$$
{\cL} \;=\; \psi^+ \, i(\partial_0 - i\,A_0)\;
\psi + \frac{1}{2m}\; \psi^+ (\partial_i - i\,A_i)^2 \psi \eqno(11.8)
$$
Here $A_\mu$ denotes the external fixed electromagnetic
potential. To attach flux to the electrons, consider the Lagrangian
$$
\eqalign{
{\cL}'\;&=\; \psi^{+'} i \bigl( \partial_0 - i(\alpha_0 - A_0)\bigr)
\psi' + \frac{1}{2m} \psi^{+'} \bigl(\partial_i - i (\alpha_i - A_i)\bigr)^2
\psi' \cr
&+\; \frac{1}{4 \pi p} \,\epsilon\, \alpha\, \partial \,\alpha\,.\cr}
\eqno(11.10)
$$
\noindent Here $\alpha_\mu$ denotes a gauge potential. From our
discussion in section I we see that we have attached $p$ units of flux to
each electron. These ``dressed" electrons are now moving in a reduced
magnetic field \
$B_{\rm{eff}}\;=\;B\;- b\;=\;B\;- 2 \pi p n\,.$

Recall  that in
Jain's argument we would like to relate our fractional Hall field to an
integer Hall fluid with filling factor $\nu=m$. The transition between the
$m$ Landau levels involve short distance high frequency physics which we
are ignoring. We thus argue that we are justified in treating the different
Landau levels as independent degrees of freedom.
We thus  replace the fermionic field $\psi'$ in \(11.10) by $m$ bosonic
fields $\phi_I, I=1,2,\dots,m$.
Recall from section II that we can change statistics by coupling
a Chern-Simons gauge potential to each bosonic field.  Thus, we argue that
the
integer Hall system may be described by the effective theory
$$
\eqalign{
{\cL}\;=\;\sum_{I=1}^m &\biggl\{ \phi_I^+\,i \bigl(
\partial_0 - i (\alpha_0 + a_{I0} - A_0)\bigr) \phi_I \;
+\;\frac{1}{2M} \phi_I^+ \bigl( \partial_i - i (\alpha_i + a_{Ii} - A_i)
\bigr)^2 \phi_I \cr
&\qquad -\; V_I (\phi_I^+ \phi_I)\;+\; \frac{1}{4 \pi}
\varepsilon a_I \partial a_I \biggr\} \;+\;
\frac{1}{4 \pi p} \varepsilon \alpha \partial \alpha\,.\cr}
\eqno(11.12)
$$
\noindent Here $\phi_I$ is a bose field corresponding to the
I$^{\rm{th}}$
Landau level, and $a_I$ is a gauge potential that converts the bose field
$\phi_I$ into effectively a fermionic degree of freedom. The potential
$V_I (\phi_I^+ \phi_I)$, whose detailed form we don't need to specify,
represent a possible short distance interaction between electrons
in the  I$^{\rm{th}}$ Landau level. We have stacked $m$ copies
of the effective theory
for the $\nu = 1$ Hall system on top of each other in order to describe the
$\nu = m$ Hall system.\refto{wniu, ETHE}

Unfortunately, at this point we have to interrupt the flow of our narrative
 in order to introduce a formalism that we need, that of duality
transform.

\head{VI. Duality and Vortices}
\taghead{6.}

In \(11.20) we have a number of scalar boson fields. It turned out that we
can represent the effective degrees of freedom contained in these scalar
fields by using gauge potentials. In this section we will describe this
so-called duality formalism,\refto{WZDD1, mpaf, WZDD2} which is of
importance in its
own
right.

Consider a relativistic Landau-Ginzburg theory
$$
\cL =   |\part_\mu \Phi|^2 - a|\Phi|^2 - b|\Phi|^4.
\eqno(LG)
$$
In the
superconducting
phase, $a$ is negative, and  we have a Goldstone mode $\eta$
corresponding
to the phase of $\Phi : \Phi=|\Phi| e^{i\eta}$.  The dynamics of $\eta$
is described by the Lagrangian
$$\cL={1\over 2h^2} (\part_\mu \eta)^2 \eqno(9.1)$$
where we have chosen spacetime units so that the phonon velocity is unity
and
we denote $|\Phi|^2$ by $h^{-2}$.
I am describing the relativistic theory here  for pedagogical reasons: the
formulas
are  much simpler and the results are essentially the same. Later in
this section, I will give a non-relativistic treatment.
(Throughout this section we will treat the electromagnetic gauge field
$A_\mu$
as a background classical field that we can switch on at the end of our
calculation.)

In (2+1)-dimensional spacetime,
we can go to a dual representation by introducing a gauge potential
$a_\mu$
and a gauge field $f_\munu =\part_\mu a_\nu -\part_\nu a_\mu$ related to
$\eta$ by
$$\part_\mu \eta = \be h^2\eps_{\munu\la} f^{\nu\la} \eqno(9.2)$$

We will now explain a choice of the arbitrary overall constant $\be$, which
sets the scale of $f_\munu$, so that  the electromagnetic current $J_\mu$
takes on a convenient form.  Had we switched on electromagnetism through
the covariant derivative $(\part_\mu -iA_\mu) \Phi$ we would have gotten,
instead of \(9.1), $\cL={1\over 2h^2} (\part_\mu \eta -A_\mu)^2$.  Thus,
we identify
the electromagnetic current (or equivalently the velocity current in the
superfluid) $J_\mu = \part_\mu \eta/h^2$ (this is standard, of course) and
hence $J_\mu =\be \eps_{\munu\la} f^{\nu\la}$ according to \(9.2).  (Note
that
we have suppressed the charge ($=2e$ for superconductors for example)
 carried by the order parameter
$\Phi$.)

In addition to the gapless excitation $\eta$, the superfluid also contains
a gapful excitation, namely the vortex, one at each place where $|\Phi|$
vanishes.  Consider the ``electric" charge coupled to the gauge potential
$a_\mu$.  Around this electric charge there is an electric field
$f_{0i}\propto x_i/r^2$.  According to \(9.2), this describes a vortex flow
$v_i\propto \eps_{ij} x_j/r^2$.  Thus, we see that vortices correspond to
electric charges in the dual theory. This fact lies  at the heart of the
duality
picture. The reason we want to go to the dual picture is that the vortex,
 a complicated soliton in the original Landau-Ginzburg
picture, is described by an elementary field in the dual picture.

We will now choose $\be$ so that the
vortex carries unit ``electric'' charge.  First, note that with
\(9.2), the Lagrangian \(9.1) in the dual picture is just good old Maxwell
$$\cL=-\be^2 h^2 f_\munu f^\munu \eqno(9.1')$$
as in \(8.4).
Thus, if we couple $a_\mu$ to a current $j_\mu$ by adding $a_\mu j^\mu$
to \(9.1')  we obtain the Maxwell equation
$$4\be^2 h^2 \part^\mu f_\munu =j_\nu .\eqno(9.3)$$
Integrating the time component of this equation over space, normalizing
the vortex charge
$$
\int d^2 xj_0=1=\int d^2x2\beta\partial_i(\eps_{ij}\partial_j\eta),
\eqno(nm)
$$
 and keeping in mind the quantization
of vortex circulation, namely $\oint dx_i \part_i \eta=2\pi$, we determine
that $\be=1/(4\pi)$ and thus
$$J_\mu ={1\over 4\pi} \eps_{\munu\la} f^{\nu\la} .\eqno(9.4)$$

In summary then, the following effective Lagrangian
$$\eqalignno{ \cL =&{1\over 4g^2} (2f_{0i}^2 -c^2 f_{ij}^2) + \vert
(\part_0
-ia_0)\phi\vert^2\cr
&+ v^2 \vert (\part_i -ia_i) \phi\vert^2 -m^2 \phi^\dagger\phi &(9.5)\cr}$$
describes the long distance dynamics of the superfluid.  We have denoted
the field describing the vortex in the dual picture by $\phi$. This is to be
supplemented by the equation for the electromagnetic current
$$J_i = {1\over 2\pi} \eps_{ij} f_{0j} .\eqno(9.6)$$
We have written a
non-relativistic
 Lagrangian in \(9.5) with the phonon velocity
$c$ and the vortex velocity $v$.  We have taken the low momentum
dispersion
of the vortex to have the form $\om=\sqrt{m^2+v^2k^2}\simeq m^2
+{v^2\over
2m^2} k^2$.  Note that $m^2$ represents the gap for creating a
vortex-antivortex pair and $m/v$ the effective ``mass'' of the vortex.

As promised, I will now give a non-relativistic treatment of duality. The
treatment given here follows essentially that given by Fisher and
Lee.\refto{mpaf} Let us start with bosons interacting via
some
short range repulsion
$$\cL=i\phi^+\part_0\phi-{1\over2m}\part_i\phi^+\part_i\phi-
\la(\phi^+\phi-
\bar\rho)^2\eqno(9.8)$$
We have written the usual quartic potential in a form emphasizing that
these
bosons are moving in a ``neutralizing" background as is the case in actual
applications to condensed matter systems.

We substitute
$$\phi=\rho^{1/2} e^{i\eta}\eqno(9.9)$$
in \(9.8) to obtain
$$\cL=-\rho\part_0\eta-{\rho\over2m}(\part_i\eta)^2-\la(\rho-
\bar\rho)^2+\dots
\eqno(9.10)$$
which we rewrite as
$$\cL=-\xi_\mu\part^\mu
\eta+{m\over2\rho}\xi^2_i-\la(\rho-\bar\rho)^2+\dots\eqno(9.11)$$
In \(9.10) we have dropped a term
$\sim(\part_i\rho^{1/2})^2$. In \(9.11) we have defined
$$\xi_0\equiv\rho\eqno(9.12)$$
Integrating out $\xi_i$ in \(9.11) we recover \(9.10).

We introduce  vortices by decomposing the angle variable into two parts:
$$
\eta=\eta_{\rm smooth} +\eta_{{\rm vortex}}\eqno(9.13)
$$
The field $\eta_{\rm smooth}$ represents the
 phonon while $\eta_{{\rm vortex}}$ is
such
that when we go around a place where a vortex is located,
$\eta_{{\rm vortex}}$ changes
by
$2\pi$. The first term in \(9.11) become $\xi_\mu\part^\mu\eta_{\rm
smooth}
+\xi_\mu\part^\mu\eta_{{\rm vortex}}$.
 Integrating out the phonon field $\eta_{\rm
smooth}$, we obtain the constraint
$$\part^\mu\xi_\mu=0\eqno(9.14)$$
which is solved by writing
$$\xi^\mu=\eps^{\mu\nu\la}\part_\nu\hat a_\la\eqno(9.15)$$
The hat \^\ on $a_\mu$ is for later convenience.

It is at this point that the gauge field comes in. Note that
$$\xi_0\equiv\rho=\eps_{ij}\part_i\hat a_j\equiv\hat f\eqno(9.16)$$
is the ``magnetic" field strength while
$$\xi_i=\eps_{ij}(\part_0\hat a_j-\part_j\hat a_0)\equiv \eps_{ij}\hat f_{0j}
\eqno(9.17)$$
is the ``electric" field strength.

The coupling to the vortex angle $\eta_{\rm vortex}$ is given by
$$\eqalign{\xi_\mu\part^\mu\eta_{\rm vortex}&=
\eps^{\mu\nu\la}\part_\nu\hat a_\la
\part_\mu\eta_{\rm vortex}\cr
&=\hat a_\la\left(\eps^{\mu\nu\la}\part_\nu\part_\mu\eta_{\rm
vortex}\right)\cr}
\eqno(9.18)$$
According to Leibniz and Newton, differentiation is a commuting
operation and
so the operator $\eps^{\mu\nu\la}\part_\nu\part_\mu$ should be zero.
Here
it is
operating, however, on a topologically singular function. The quantity in
parenthesis is non-zero and
in fact equal to
the vortex current $j^\la_{\rm vortex}$ up to a factor
of $2\pi$. To see this we integrate $j^0_{\rm vortex}$ over a circle
containing
a vortex. Using Stoke's theorem, we find (this is of course the same as
\(nm))
$$\eqalign{\int d^2x\  j^0_{\rm vortex}&=\int
d^2x\ \eps_{ij}\part_i\part_j\eta_{\rm vortex}
=\oint dl_j\ \part_j\eta_{\rm vortex}\cr
&=2\pi\cr}\eqno(9.19)$$
Henceforth, we will ignore factors of $2\pi$ \etc in the interest of keeping
the
equations simple.

Putting all of this into \(9.11) we have
$$\cL=\hat a_\mu j^\mu_{\rm vortex} +{m\over\rho}\hat f^2_{0i}-\la(\hat
f-\bar\rho)^2+\dots\eqno(9.20)$$

To ``subtract out" the background $\bar\rho$, an obviously sensible move
is to
write
$$\hat a_\mu=\bar a_\mu+a_\mu\eqno(9.21)$$
where we define the background gauge potential by $\bar a_0=0$ and
$$\eps_{ij}\part_i\bar a_j=\bar\rho\eqno(9.22)$$
The Lagrangian \(9.20) then has the cleaner form
$$\cL=\left({m\over\bar\rho}f^2_{0i} -\la f^2\right)+a_\mu j_\mu^{\rm
vortex}\ +\bar a_i j^{\rm vortex}_i\eqno(9.23)$$
We have expanded $\rho\approx\bar\rho$ in the first term. The theory
contains a
propagating mode with speed
$$c=(\la\bar\rho/m)^{1/2}\eqno(9.24)$$

In suitable units in which $c=1$, we now have the Maxwell Lagrangian
$$\cL=-{1\over2g^2}f^2_{\mu\nu}+a_\mu j_\mu^{\rm vortex}+\bar a_i
j_i^{\rm
vortex}\eqno(9.25)$$
Compare this with \(9.5). Interestingly, the spatial part $f_{ij}^2$ of the
Maxwell Lagrangian comes from the short ranged repulsion between the
original
bosons.

The one thing we missed with our relativistic treatment is the last
term in \(9.25), for the simple reason that we didn't put in a background.
Recall that a term like $A_iJ_i$ in ordinary electromagnetism means that a
moving particle associated with the current $J_i$ sees a magnetic field
$\vec
\nabla\times\vec A$. Thus, a moving vortex will see a ``magnetic field"
equal to
the sum of the density of the original bosons and a fluctuating field
$$\eps_{ij}\part_i(\bar a+a)_j=\bar\rho+\eps_{ij}\part_ia_j\eqno(9.26)$$

In  the Coulomb gauge
$$\part_ia_i=0\eqno(9.27)$$
we have
$$(f_{0i})^2=(\part_0a_i)^2+(\part_i a_0)^2\eqno(9.28)$$
where the cross term $(\part_0 a_i)\ (\part_i a_0)$ vanishes upon
integration by
parts.  Integrating out the Coulomb field $a_0$ we have
the
Lagrangian as written in the condensed matter literature
$$\eqalign{\cL&={1\over2g^2}\left[(\part_0 a_i)^2-(\part_ia_j-
\part_ja_i)^2
\right]\cr
&+g^2\int\int d^2xd^2y\ j^{\rm vortex}_0(x)\log |x-y|j_0^{\rm
vortex}(y)\cr
&+(a_i+\bar a_i)j_i^{\rm vortex}\cr}\eqno(superfluid)$$

If we had taken the bosons to interact by an arbitrary potential $V(x)$ we
would have, instead of the last term in \(9.11),
$$\int\int d^2xd^2y\left(\rho(x)-\bar\rho\right)
V(x-y)(\rho(y)-\bar\rho)\eqno(9.30)$$
It is easy to see that all the steps go through essentially as before, but now
the second term in \(9.23) becomes
$$\int\int d^2xd^2y f(x)V(x-y)f(y)\eqno(9.31)$$
Thus, the gauge field propagates according to the dispersion relation
$$\omega^2=(\bar\rho/m)V(k)\vec k^2\eqno(9.32)$$
where $V(k)$ is the Fourier transformation of $V(x)$.
In the special case $V(x)=\la\de^{(2)}(x)$ we recover the linear
dispersion given in \(9.24). Indeed, we have a linear dispersion $\omega\,\,
\propto\ \, |\vec k|$
as long as $V(x)$ is sufficiently short ranged for $V(\vec k=0)$ to be
finite.

An interesting case is when $V(x)$ is logarithmic. Then $V(k)$ goes
like $1/k^2$ and so $\omega\sim$ constant:
 the gauge field $a_i$ becomes massive. Referring to \(superfluid)
we see that a theory of bosons with a logarithmic repulsion between them is
in
the long distance approximation self dual.

Having constructed this duality formalism, let us reward ourselves by
deriving the motion of vortices in a fluid. Let the bulk of the fluid be at
rest. According to \(superfluid) the vortex behaves like a charged particle
in a background magnetic field $\bar b$ proportional to the mean density
of the fluid $\bar \rho$. Thus, the force acting on a vortex is  the usual
Lorentz force $\vec v \times \vec B$, and so the equation of motion of the
vortex in
the presence of a force $F$ is just
$$
\bar \rho \epsilon_{ij} \dot x_j = F_i
\eqno(ext)
$$
This is the well-known result that a vortex, when pushed, moves in a
direction perpendicular to the force.

Consider two vortices. According to \(superfluid) they repel each other by
a logarithmic interaction. They move perpendicular to the force. Thus,
they end up circling each other. In contrast, consider a vortex and an
anti-vortex.
They attract each other. As a result of this attraction, the vortex
and the anti-vortex both move in the same direction, perpendicular to the
straight line joining them. The vortex and anti-vortex move along in step,
maintaining the distance between them. This in fact accounts for the famous
motion of a smoke ring. If we cut a smoke ring through the center of the
ring and perpendicular to the plane the ring lies in, we have just a vortex
with an anti-vortex for each section.
 Thus, the entire smoke ring moves along in a direction
perpendicular to the plane it lies in.

All of this can be understood by elementary physics, as it should be. The
key observation is simply that vortices and anti-vortices produce circular
flows in the fluid around them, say clockwise for vortices and
anti-clockwise
 for anti-vortices. Another basic observation is that if there is a
local flow in the fluid, then any object, be it a vortex or an anti-vortex,
caught in the flow would just flow along in the same direction as the local
flow. This is a consequence of Galilean invariance. The reader can
convince himeslf or herself by drawing a simple picture that this produces
the
same  pattern of motion as discussed above.

This discussion of duality and vortices can be generalized readily to
(3+1) dimensional spacetime.\refto{vortex}

This discussion of duality is not quite rigorous. From standard topological
discussions, we know that at the location of the vortex the magnitude field
$\rho$ must vanish. In other words, $\eta_{\rm vortex}$ and $\rho$
should be
coupled. This
was ignored in the derivation. To give a rigorous discussion, we would
have to put the theory on a lattice.

Let us also sketch what happens when the electromagnetic potential is
turned
on. We replace ordinary derivative $\part_\mu$ by covariant derivative
$\part_\mu-iA_\mu$. In \(9.11) instead of $\xi^\mu\part_\mu\eta$ we
have
$\xi^\mu (\part_\mu\eta - A_\mu)$. We thus have to drag along an extra
term $\xi A$ and when $\xi$ is replaced by the curl of a gauge potential
this term becomes
$A_\mu\eps_{\mu\nu\la}\part_\nu\hat
a_\la=A_0\bar\rho+A_\mu\eps_{\mu\nu\la}
\part_\nu a_\la$. The $A_0\bar\rho$ term is cancelled by the coupling of
$A_0$
to the background.
We see that the only modification in the Lagrangian in \(9.25) is that the
vortex current is replaced by ($j^{\rm vortex}_\mu - \eps_{\mu\nu\la}
\part^\nu
A^\la$). When we integrate out $a_\mu$ (and suppressing $j_\mu^{\rm
vortex}$
for simplicity),
 we have schematically
$$\eqalign{ \cL &\sim a\part^2a+a\eps\part A\cr
&\sim \eps\part A\ {1\over\part^2}\ \eps\part A\cr
&\sim A^2\cr}\eqno(meiss). $$
We have found
the Meissner effect and hence superconductivity.

Compare this with \(12.5) or \(11.20) we see how the Meissner effect and
the Hall effect are intimately related. It is merely a question of how many
derivatives act on little $a$ and big $A$. We summarize with a table:
\bigskip

\begintable

$a \part^2 a$ | $A^2$| Meissner \cr

$a \eps\part a$ |$ A \eps\part A$| Hall \cr

$a^2$ | $A \part^2 A$| Maxwell \endtable

For completeness, we have included a third possibility:  if for some reason
the gauge
potential acquires a gauge symmetry breaking mass, we would have
$$\eqalign{\cL&\sim a^2+a\eps\part A\cr
&\sim \eps\part A \eps\part A\cr
&\sim F^2}\eqno(8.17)$$
We obtain the Maxwell term.

In a sense, this is duality at work\refto{WZDD1}.

Also, some of you may remember from Wiegmann's lectures that he
considered superconductivity in (1+1) dimensions. Instead of the
(2+1) dimensional coupling $A_\mu \epsilon^{\mu\nu\la} \part_\nu
a_\la$ we have its (1+1) analog   $A_\mu \epsilon^{\mu\nu} \part_\nu
\phi$. Thus, if  $\phi$ obeys dynamics governed by $(\part \phi)^2$, then
we obtain the Meissner effect and superconductivity in exactly the same
way as in \(meiss).

Incidentally, here we have also cleared up a potential source of confusion
in
the discussion following \(9.2). For the Lagrangian
$\cL={1\over2h^2}(\part_\mu\eta-A_\mu)^2$, we were rather sloppy in
identifying
the current to be
$J_\mu=\part_\mu\eta/h^2=\be\eps_{\mu\nu\la}f^{\nu\la}$. Some
people may be confused by the quadratic $A_\mu^2$ term in the
Lagrangian and
wonder if the current $J_\mu$ should depend on $A_\mu$. (The subtlety
here
hinges on the gauge invariance of $J_\mu$.) The procedure given here
provides a
resolution of this point. Let us then follow the steps outlined above and
which, at the risk of repeating ourselves, we will sketch here. Starting with
$\cL\sim(\part_\mu\eta-A_\mu)^2\sim\xi_\mu(\part_\mu\eta-
A_\mu)+\xi^2_\mu$ we
integrate out $\eta$ to obtain the constraint $\part^\mu\xi_\mu=0$ which
we
solve by writing $\xi_\mu=\eps_{\mu\nu\la}\part_\nu a_\la$. Plugging in,
we
find $\cL\sim f^2+A_\mu \eps_{\mu\nu\la}\part_\nu a_\la$. The current is
indeed
as given above. (All of this is to done at the level of the path integral of
course and may be made completely rigorous by latticizing spacetime.)

\head{VII. Multitude of Gauge Fields}
\taghead{7.}

After this digression on duality, the reader may have forgotten where we
were. Let me remind you that we had reached an effective Lagrangian
\(11.12)
containing a bunch of complex scalar fields $\phi_I$. Each of these fields
contains a  gapless Nambu-Goldstone phase degree of freedom  and a
gapful vortex degree of freedom. We now use the  dual
representation to describe the physics directly in terms of these
physical degrees of freedom rather than the complex scalar fields $\phi_I$.

We can thus write the Lagrangian \(11.12) in the dual representation as
$$
\eqalign{
{\cL}\;=\;\sum_{I=1}^m\;\biggl\{ \frac{2}{4 \pi}
&(\alpha + a_I - A)\, \varepsilon\partial \alpha_I\;+\;
\frac{1}{4 \pi}  \,a_I\,\varepsilon\partial\,a_I \;+\;
\alpha_{I\mu}\, j_{I\;\rm{vortex}}^\mu \biggr\}\;+ \cr
&\frac{1}{4 \pi p}  \alpha \varepsilon \partial \alpha \;+\;\cdot\,\cdot\,
\cdot\;.\cr}
\eqno(11.13)
$$
\noindent We have introduced $m$ gauge potentials $\alpha_{I\mu}$,
one for each of the Nambu-Goldstone mode contained in the $m$ complex
scalar fields $\phi_I$. The vortex degrees of freedom are contained in
$j_{I\mu}$, the
I$^{\rm{th}}$ vortex current to which $\alpha_\mu$ couples.
Notice that we have also not explicitly displayed the Maxwell terms
$f_{I \mu \nu}^2$ for the gauge potentials $\alpha_{I \mu}$.
Compared to the various Chern-Simons terms we displayed, the Maxwell
term
has one power of mass dimension higher and so is unimportant at long
distances. Our effective theory does not describe the short distance
physics in any case.

The long distance physics is now described entirely
in terms of gauge fields, of which we have introduced a multitude. It may
be helpful to recall how they entered. The gauge potential $\alpha$
attached flux to the electrons. The gauge potentials $a_I$ changed the fermi
statistics of the degrees of freedom contained in the Landau levels to
bosonic statistics. Finally, the gauge potentials $\alpha_I$ embodied the
phase degrees of freedom contained in the bose fields representing the
Landau
levels.

We see that in \(11.13) only the gauge potentials $\alpha_I$ couple to the
quasi-particles (the vortices) and to the external electromagnetic potential.
Thus, to determine the response of the system to an external
electromagnetic probe and the quantum numbers of the quasi-particles we
might as well
integrate out $a_I$ and $\alpha$.  Note that the Chern-Simons term is
``self-reproducing''.  Schematically, as we have seen already, if we have
$\cL\sim a\eps\part a+a\eps
\part b$, then integrating out $a$, we would get $\cL\sim (\eps\part b)
(\eps\part)^{-1} (\eps\part b)\sim b\eps\part b$ by simple cancellation.
This can be made precise
with the formulas given in section II:
integrating out $a$ in
$$
\cL= {1\over 4\pi} a \epsilon \part a +  {1\over 2\pi} a \epsilon \part b
\eqno(pre1)
$$ we obtain
$$
 \cL= {1\over 4\pi} b \epsilon \part b
\eqno(pre2).
$$
Basically,  we are just saying
that  the Chern-Simons term is the unique gauge and
scale
invariant term we can write down.

Proceeding thus, we obtain, after integrating out $a_I$,
$$
{\cL}\;=\;\frac{1}{4\pi}\,\biggl[\sum_I \bigl\{
2  (\alpha - A)\varepsilon \partial \alpha_I +
 \alpha_I\varepsilon \partial \alpha_I\bigr\} \;
+\; \frac 1 p \varepsilon \alpha \partial \alpha \biggr]\;+\;
\sum_I \alpha_I j_I^{\rm{vortex}}\,. \eqno(11.14)
$$
\noindent Next we integrate out the gauge potential $\alpha$ to find
$$
{\cL}\;=\;\frac{1}{4\pi} \biggl[ \sum_I  \alpha_I\varepsilon
\partial \alpha_I \,+\,p \sum_{IJ} \alpha_I\varepsilon \partial \alpha_J
\biggr] \;
+\; \frac{2}{4 \pi} \sum_I \biggl( -
\alpha_I\varepsilon \partial A \,+\, 2
\pi \alpha_I \;j_I^{\rm{vortex}}\biggr)\,.
\eqno(11.15)
$$

We see that we have obtained precisely the effective Lagrangian \(cop) we
obtained
earlier with our ``slick" approach based entirely on general principles. The
only difference is that in this derivation $p$ ``automatically" comes out to
be an even integer. Recall that in the ``slick" approach, we fix $p$ to be an
even integer by requiring that  the electron be  a fermion. In this
approach $p$ is fixed to be an even integer from its introduction but the
underlying physics is essentially the same, namely quantum statistics.

I would like to end this
section by remarking that the effective Lagrangians
written down by
different authors often look quite different. The reason is
simply that
different authors choose to integrate out different combinations
of
the
gauge potentials. For illustration, let us derive the Lagrangian first
written
down by Zhang \etal\refto{ZHG} for filling factor $\nu=1/(p+1)$
with
$p$ even.

Go to \(11.12) and specialize to $m=1$. We
have
$$
\cL=\phi^+i\left(\part_0-i(\a+a-
A)_0\right)\phi+(\dots)
+{1\over4\pi}\left(\eps a\part a+{1\over
p}\eps\a\part\a\right)
\eqno(11.40)
$$
(To avoid clutter we have
represented the term with the spatial covariant
derivative acting on $\phi$
by (\dots).) It turns out that to obtain the
Lagrangian of Zhang \etal\refto{ZHG} we
should integrate out the combination $(a-
\a)$.
Let us write
$a=(\ga+\be)/2$ and $\a=(\ga-\be)/2$. The last two terms in
\(11.40)
became
$${1\over4}\cdot{1\over4\pi}\left[{(p+1)\over p}
\left(\eps\ga\part\ga+\eps\be
\part\be\right)+{2(1-p)\over
p}\eps\a\part\be\right]\eqno(11.41)$$
Integrating out $\be$ we
obtain
$$
\cL=\phi^+i\left(\part_0-i(\ga-
A)_0\right)\phi+(\dots)
+{1\over4\pi}{1\over(p+1)}\eps\ga\part\ga\eqno(11.42)
$$
This is the original Lagrangian of Zhang \etal \ The reader would
realize
that
this Lagrangian may be written down immediately: the
Chern-Simons term
for $p$
even simply changes the statistics of $\phi$ to
fermionic. The point is that
the more
elaborate Lagrangian in \(11.12)
allows us to describe more complicated
filling
fractions.

\head {VIII. Topological Order}
\taghead{8.}

As I mentioned in the Introduction the Hall fluid is an example of a
topological quantum fluid, but so far I have not really mentioned topology.
What is topological about the Hall fluid? As I will now explain, the Hall
fluid is an example of strongly correlated electron systems
in two spatial dimensions, and one fascinating feature
of such systems is that they may have ``global'' and ``topological''
properties
not reachable, or at least not easily reachable, with conventional condensed
matter physics methods.
The long distance physics of
the ground state  and its low lying
excitations is described by a new class of
field theories known as topological field theories.  Topological field
theories were first investigated for
their possible relevance to string theory.\refto{TF}
Thus, they may well be field theories in
search of physical realization.  It is amusing then that at least some
watered-down
version may be realized in condensed matter systems.

Let me first remind you of the notion of order. A standard example in
condensed matter physics is the ferromagnet. As is familiar, there is a  long
range order in the ferromagnet: all the spins are lined up in the same
direction. The long range order is intimately connected with spontaneous
symmetry breaking. In the ferromagnet, rotational invariance is
spontaneously broken. Thus, the order in condensed matter physics, in
particle physics, and other areas of physics, may be classified by symmetry
groups. All this is familiar textbook material.

In stark contrast, there is no
identifiable broken symmetry in the quantum Hall fluid and the order is not
characterized by some symmetry group being broken. Instead,
Wen\refto{WTF} has
emphasized that these systems enjoy a new type of order called
``topological order," which he has described picturesquely as a ``dancing
pattern" that the electrons in the fluid are obliged to follow because of the
twin dictates of Fermi statistics and the Lorentz force from the external
magnetic field, as I explained in Section~I. Topological order is an
entirely new concept in condensed matter physics. It is not characterized by
a symmetry group. Rather, it is characterized by the $K$ matrix. The $K$
matrix specifies the dance the electrons are to follow. With different $K$
matrices, the dancing electrons trace out a different pattern.

So what is topological? Let us look at the Chern-Simons action, for
simplicity written for  a single gauge
potential $a$:
$$
S=\int d^3 x   {k\over 4\pi}\epsilon^{\mu\nu\lambda}
a_\mu\partial_\nu a_\lambda.
\eqno(top)
$$

There is something rather unusual about this action. When we learned
about field theory we first encountered the free scalar field theory
described by the action
$$
S = \int d^3 x   {\sqrt g}(g^{\mu\nu} \part_\mu \phi \part_\nu \phi - m^2
\phi^2)
\eqno(scalar)
$$
We need the metric $g_{\mu\nu}$ to
contract the Lorentz indices in the kinetic energy term. Furthermore, the
determinant of the metric $g$ is needed to produce the correct volume of
spacetime. This is of course as it should be: the metric tells us about
distances in spacetime. To do physics, we have to have rulers and clocks.
We can hardly have a more basic statement about physics than
that. Usually, of course, the metric is suppressed since normally we are
interested only in field theory in flat spacetime. Nevertheless, in the back
of our minds, we must remember that the metric is always there. It is just
normally set to be equal to the Lorentz metric and then suppressed.

But in \(top) the metric does not come in at all. The indices are contracted
with  the antisymmetric  symbol
$\epsilon^{\mu\nu\lambda}$,
not with the metric
$g^{\mu\nu}$. The reader should also verify that the action in \(top) is
perfectly covariant as written. In particular, no factor of ${\sqrt g}$ is
needed. In other words, the action is topological.
We have entered an unfamiliar realm of physics: a realm without rulers
and
without clocks.

By way of contrast,
the Maxwell term cannot be written without the metric
$g_{\munu}$.
Thus, with the Maxwell term
the action takes the form
$$S=\int d^3x \left[
{k\over 4\pi} \eps^{\munu\la} a_\mu \part_\nu a_\la
-\sqrt
g {g^{\mu\la}
g^{\nu\si}\over 2g^2} f_\munu f_{\la\si}
+\ldots\right]\eqno(11.2)$$
(I trust the reader not to confuse the gauge coupling
$g$ with the
determinant of the
metric, and I have written, for compactness of notation, the Maxwell term
in its Lorentz invariant form.)
The pure Chern-Simons theory depends only
on the topology of spacetime,  not on the metric: it is a
topological field theory of the type
first studied by Witten.\refto{TF}

This is a powerful
statement: it suggests those properties of the
microscopic theory that
depend on the metric cannot affect the Chern-Simons term.
Heuristically, we can argue that to describe disorder and impurities we
necessarily have to drag in the metric, for example to measure the energy
of the potential well associated with an impurity with. As another example,
we can suppose that at short distances electrons in the actual sample move
on a lattice. An imperfection in the lattice may mean that the ability of an
electron to hop from site to site is affected by the imperfection. This may
be modeled by saying that distance and time are distorted near the
imperfection. Thus, we argue that
disorder and impurities cannot affect the Chern-Simons term and hence the
long
distance physics of the Hall
fluid, such as the charge and statistics of the quasiparticles in the fluid, as
determined in section IV.

Having heard all this, you may start to wonder. You say, wait a minute!
Isn't the stress energy tensor defined as the
variation
of the action with respect
to
$g^{\mu\nu}$? But since the action does not depend on the metric, the
stress energy tensor, from which physical quantities like energy and
momentum are derived,  is identically zero! In fact, this is as it should be.
After
all, the concepts of energy and
momentum depend on rulers and clocks.
In particular, the Hamiltonian is exceedingly
simple:
it is just zero.

In school, when we took a course on quantum mechanics, we did plenty of
homework problems in which we were given a Hamiltonian and told to
determine its energy spectrum. Boy, it would have been a nice simple
problem had the professor given us a problem in which the Hamiltonian is
identically zero! The energy spectrum is easy to determine, then.
All states have
zero energy. (More precisely, the scale of energy (or mass) is
set by the Maxwell
coupling
$g^2$, as we saw in section II. In a purely topological theory, the absence
of the non-topological Maxwell term means that effectively $g^2$ has been
sent to $\infty$.)

But to get full credit, it is not enough to just say that all states have zero
energy. We have to determine
how many such zero energy states there are, as was first emphasized by
Witten. The number is known as the ground state degeneracy. Let us study
the Chern-Simons theory on a closed two-dimensional surface of genus G.
(The genus simply measures the number of holes or handles. Thus, the
sphere has genus 0, the torus genus 1, and so on.) This is not as far fetched
as it may sound: if we impose periodic boundary conditions on a system
defined on a rectangle, we are effectively dealing with a system defined on
a torus. We will consider surfaces with boundaries later.

The ground state
degeneracy can only depend on the
topology of the closed
two-dimensional surface. Let us calculate it for a torus, say.  We will
follow
the discussion given by Wen.\refto{WTF}
The basic idea goes back to an
observation
by Wu and Zee\refto{WUZ}
 that a topological term in field
theory may be regarded as due to
the presence of a gauge field in field
configuration space. This is a real
mouthful but is easy to understand.
Suppose we are given a Lagrangian
describing the motion of a particle
with coordinates $q_i$.  If in the
Lagrangian we find a term linear in the
time derivative $\dot q_i$, then we
would know that the particle is moving
under the influence of an
electromagnetic
potential $\cA_i$, where
$\cA_i$ is just the coefficient of $\dot q_i$.

We have precisely this
situation in a Chern-Simons theory.  In the $a_0=0$
gauge  the Lagrangian
is
$$\cL= {k\over4\pi} \eps_{ij} a_i\dot a_j\eqno(11.3)$$
Thus, we see
that in the configuration space of the gauge potentials, a
magnetic field is
present.  The analog of Gauss' law, obtained by varying
with
respect to
$a_0$, tells us that
$$\eps_{ij} \part_i a_j=0\eqno(11.4)$$
This tells us
that $a_i=a_i(t)$ in the gauge $\part_ia_i=0$.  In the $a_0=0$
gauge, we
can still make time  independent gauge transformations
$a_i\to a_i-iu^{-1}\partial_i u$ with $u(x)$ independent of time.
  On a torus, the
allowed gauge transformation must have the form
$$
u=e^{i2\pi({n_1x^1\over L_1}+{n_2x^2\over L_2})}
\eqno(11.5)
$$
where $n_1$ and $n_2$ are integers,  since
$u$ cannot depend on whether we
coordinatize the same point as
$(0,x^2)$ or $(L_1,x^2)$, for example.
The
gauge potential $a'_i=a_i-
iu^{-1} \part_iu= a_i+2\pi n_i/L_i$ is equivalent
to
$a_i$.  Purely for the
sake of convenience, let $a_i(t)=(2\pi/L_i)q_i(t)$.
We have just learned
that $q_i$ and $q_i+1$ are equivalent.  In other words, the dynamics
may now be described in terms of a point particle living on a torus of
length 1 and width 1.

Plugging in, we find the action
becomes
$$\eqalign{ S=&\int d^3x{k\over 4\pi} \eps^{\munu\la} a_\mu
\part_\nu
a_\la\cr
=&k{L_1L_2\over 4\pi} \int dt {(2\pi)^2\over
L_1L_2} \eps_{ij} q_i\dot
q_j\cr}\eqno(11.6)$$
It is convenient to
regulate the theory by adding a kinetic energy term:
$$S=\int dt (k\pi
\eps_{ij} q_i\dot q_j+\12 m\dot q_i^2)\eqno(11.7)$$
We can let the mass
$m$ go to zero at the end.  Indeed, if we had included
the
Maxwell term
in the field theory, it would generate this kinetic energy
term
with
$m\sim 1/g^2$ ($f_\munu^2\sim \dot a_i^2\sim \dot q_i^2$).
As
$g^2\to\infty$, the Maxwell term disappears and $m\to 0$.

We have
a point  particle moving on a torus  and coupled to a gauge potential
$A_i=k\pi\epsilon_{ij}qj$, with the corresponding
 magnetic field
$B=\eps_{ij}
\part_i\eps_{lj} q_l(k\pi) =2\pi k$.  The total flux going
through the
surface
of the torus is $\Phi=BA=2\pi k$  since the area is
normalized to 1.  This
quantum mechanical problem can be solved
explicitly\refto{HR} and the
result is
that the ground state is $k$-fold
degenerate.  The degeneracy does not
depend
on $m$.

The degeneracy
can also be seen heuristically by arguing that the torus is
locally flat, but
then  we know that
in the planar problem the  the degeneracy of
each
Landau level is precisely $BA/2\pi=k$.

It can be shown that on a
surface of genus $G$ the ground state degeneracy
is
equal to $k^G$.
Heuristically, this may be argued by topologically
deforming
the surface
into $G$ ``quasi-torus'' connected by long tubes.
In effect, the dynamics of
the gauge potential $a_i(t)$ on such a surface is
the dynamics of $G$
independent particles, one on each ``quasi-torus''.
The
degeneracy is thus
$k^G$.

To an ``old-fashioned'' field theorist, the parameters in a  field theory, such
as the electric charge or the mass of the electron in  quantum
electrodynamics,
could be varied continuously, without affecting  the qualitative behavior of
the theory, at least within a certain  range. Here we really have a new type
of
field theory.  In particular, the  degeneracy formula $D=k^G$ indicates
that
the Chern-Simons theory  with $k$ not  equal to an integer cannot be
defined on
a compact surface.

In section IV we gave a physical reason why the coefficient of  Chern-
Simons
term must be equal to an integer. It is satisfying that here we have  found a
mathematical reason behind the same fact.

For further details about topological order, see the lectures by Wen in this
volume.

\head{IX. On the Edge}
\taghead{9.}

Experimentally, the Hall fluid is contained of course in a finite size region,
of the topology of a disk. Halperin\refto{HP} pointed out some years ago
that currents would flow on the edge of the disk. Physically, the origin of
these edge currents can be understood as follows.

To describe a Hall fluid confined to a disk-shaped region, we add to the
Hamiltonian a potential energy term which is essentially equal to zero
within
the disk and which rises rapidly and linearly as it crosses the boundary of
the disk. For the purpose of this discussion, we suppose that the length
 scale over which the potential varies to be much larger than the magnetic
length, namely the length defined in \(area). (For real materials, this may
not be the case.) Within the disk-
shaped
 region to which the bulk of the Hall fluid is confined, the Landau levels
are as before, but near the boundary, the Landau levels are pushed up by
the
potential energy. In general,  a  given Landau level would  cross the
 Fermi surface linearly. Thus, a
physicist ``living on the edge'' would see a linearly dispersing, that is, a
massless excitation.  Also, since there is only one branch in the dispersion,
the excitation is chiral and travels in only one direction. Thus, there must
currents flowing along the edge of the experimental sample. See the
lectures
by Wen for more details.

Within the context of our lectures, we are interested in asking whether  the
Chern-Simons term is smart enough to know about these edge currents?
Indeed, it
is.  As we have remarked in section II,
 the Chern-Simons term is not, strictly speaking, gauge invariant.
Under a
gauge transformation it
changes by a boundary term  see \(2.3):
$$
\de \int d^2 x dt a\epsilon\part a =\int du dt \Lambda (\part_0 a_u -
\part_u
a_0).\eqno(12.12)
$$
Here $u$ denotes the length coordinate along the edge. This apparent
``defect'' of the theory actually represents one
of its virtues.
To have a gauge invariant theory we must impose
the condition that $\Lambda$
must vanish on the edge. Normally, in the (1+1) dimensional world of
the edge, if we given a gauge potential $a_u$ of the form $\part_u \varphi$
we would dismiss it as an unphysical gauge degree of freedom. But here
because  of
the preceding condition on $\Lambda$ we no longer have the gauge
transformation available to gauge this degree of freedom away. Thus, the
degree of freedom embodied in $\varphi$ survives. It is this edge degree of
freedom which leads to the  edge currents.\refto{WTF,FKT}

Indeed, the dynamics on the edge world may be described by the (1+1)-
dimensional
action
$$\int dt du\12 \left[ (\part_0\varphi)^2-v^2
(\part_u\varphi)^2\right]\eqno(12.13)$$
supplemented by the constraint
$$\part_0\varphi=v \part_u \varphi\eqno(12.14)$$
Here $v$ determines the velocity of propagation of the edge degree of
freedom.

The reader who works out the derivation of \(12.14) will notice that $v$
appears as a parameter not determined by the long distance topological
Chern-Simons theory. Does this mean that the theory is stupid? In fact, the
theory doesn't, and shouldn't, know about $v$ since it is a property of
the
confining potential on the edge rather than the fluid in the bulk. Go back to
the physical derivation given at the beginning of this section. The slope of
the linear dispersion of the massless edge excitation is evidently controlled
by the confining potential.

The constraint \(12.14) means that the edge excitations
propagate in only one direction. We have thus reproduced using more
formal
methods what we had just argued on physical grounds.

  The action \(12.13), describing a massless chiral
scalar field, is the simplest example of a conformal field theory.
The theory
 has a conserved current $J^\mu=\eps^\munu\part_\nu
\varphi$ (in spacetime units such that $v$ is equal to one). Here
$\mu,\nu=0,1$ denote the (1+1)-dimensional spacetime indices. The
commutation relation between the currents is thus determined by the
commuation relation satisfied by $\varphi$ and its derivatives. We find that
the
Fourier
transforms $J_n$
of $J_0$ (assuming that the edge is bounded, such as the edge of a disk)
generate a
Kac-Moody algebra defined by
$$[J_m,J_n]= m\de_{m,-n} .\eqno(12.15)$$
In the more general case in which we have the matrix Chern-Simons theory
\(11.17)
 the
Kac-Moody algebra in \(12.15) is obviously generalized to
$$[J_m^I,J_n^J]=m\de_{m,-n} K^{IJ} .\eqno(12.16)$$
We cannot simply diagonalize the matrix $K$ because the gauge potentials
$\a_I$
couple to the vortex current $j_I^{\rm vortex}$ and the vorticity defined
by
$\int d^2x\ j_I^{\rm vortex}$ is quantized by topological considerations to
be
integers.  Thus, we are allowed to transform $K$ only to $X^TKX$ where
$X$ is
an element of $SL(m,Z)$, that is, a matrix whose entries are all integers.

Remarkably, with
$$X=\pmatrix{ 1& 0 & \ldots &\ldots & 0\cr
-1 & 1 & & & \vdots\cr
0 & -1 & 1 & & \vdots\cr
\vdots & & & \ddots & \vdots\cr
0 & \ldots & \ldots & -1 & 1\cr}\eqno(12.17)$$
we find
$$X^TKX = \pmatrix{ 2 & -1 & 0 & \ldots & \ldots & 0\cr
-1 & 2 & -1 &  & & \vdots\cr
0 & -1 & 2 & & & \vdots \cr
\vdots & & & \ddots & & 0 \cr
\vdots & & & & 2 & -1\cr
0 & \ldots & \ldots & 0 & -1 & p+1\cr}\eqno(12.18)$$

Lo and behold (once again), if we knock off the last row and the last
column we
obtain the $(m-1)$ by $(m-1)$ matrix
$$\pmatrix{2 & -1 & 0 & \ldots & 0\cr
-1 & 2 & -1 & & \vdots \cr
0 & -1 & 2 & & 0\cr
\vdots &  & & \ddots & -1\cr
0 & \ldots & 0 & -1 & 2\cr}\eqno(12.19)$$
which we recognize as the Cartan matrix for the Lie algebra $SU(n)$!
Recall
that the Cartan matrix is defined by the scalar products of the basic roots of
$SU(m)$
$$\tilde K_{IJ} = \a^I\cdot \a^J/2 \eqno(12.20)$$
for $I,J=1,\ldots, m-1$.  This remarkable fact has also been noticed
independently by Blok and Wen,\refto{bw} and by Read.\refto{NR}

Somehow the apparently simple problem of electrons moving in a magnetic
field
contains within it a non-abelian symmetry.

The transformation \(12.17) corresponds to going to a
new basis defined by $\tilde \a_1=\a_1-\a_2$, $\tilde \a_2=\a_2-\a_3$,
$\tilde
\a_{m-1} =\a_{m-1}-\a_m$, and $\tilde \a_m=\a_m$.  Physically, the
gauge
potentials $\tilde \a_I$, $I=1,\ldots, m-1$, are electromagnetically neutral
in
contrast to $\tilde\a_m$.  The $SU(m)$ symmetry transforms these $m-1$
gauge
potentials into each other.

\head {X. More General $K$ Matrices}
\taghead{10.}

We can go on and construct more general $K$ matrices.  We follow the
simple physical idea described in section IV.  Given a bunch of Hall
fluids, we can put them together and let them interact with Chern-Simons
dynamics.

Let us generalize \(12.9) and take each of the
incompressible fluids to be described by
$$4\pi \cL_I= k_I \a_I \eps\part \a_I .\eqno(12.21)$$
Then, when we put them together we obtain
$$\eqalign{ 4\pi\cL& = \sum_I k_I\a_I \eps\part \a_I + p(\sum_I\a_I)
\eps\part (\sum_J \a_J) \cr
&= \a K \eps \part \a \cr} \eqno(12.22)$$
with the matrix
$$K=\pmatrix{ p+k_1 & p & \ldots & p\cr
p & p+k_2 & & \vdots\cr
\vdots &  & \ddots & \vdots \cr
p & \ldots & \ldots & p+k_m\cr} \eqno(12.23)$$
Applying the formulas \(11.22), \(11.24), and \(11.26),
 we determine the filling factor  and conductance to
be
$$\nu=\si_H = {1\over p+{1\over \sum_I{1\over k_I}}} ,\eqno(12.24)$$
the charge to be
$$q= {1\over 1+tp} \sum_I l_I/k_I ,\eqno(12.25)$$
and the statistics to be
$${\th\over\pi} = \sum_I l_I^2/k_I - p(1+tp) q^2.\eqno(12.26)$$
Here we have defined $t=\sum_I k_I^{-1}$.
The filling factors reachable with \(12.24) include 2/3, 2/7, 8/15, 8/31, and
so
on.

We can keep on going and take for each of the component incompressible
fluids
we want to put together to be the incompressible fluid we have just
obtained.
We then arrive at
$$
4\pi \cL =\sum_S \a^S K_S \eps \part \a^S +  + 2 a\eps \part\sum_S \a^S
-{1\over p} a\eps\part a \eqno(12.27)
$$
where $K_S$ denotes an $m_S$ by $m_S$ matrix (as in \(12.23))
$$
K_S= \pmatrix{ p_S+k_1^S & p_S & \ldots & p_S\cr
p_S & p_S+k_2^S & & \vdots\cr
\vdots & & \ddots & \vdots\cr
p_S & \ldots &\ldots & p_S+k_m^S\cr}
\eqno(12.28)
$$
A short computation using \(11.22) shows that the filling factor is
$$\nu=\si_H = {1\over p+{1\over\sum_Sp_S +{1\over\sum_{I_S} {1\over
k_{I_S}^S}}}} \eqno(12.29)$$
It is intriguing that a continued fraction emerges naturally.  Clearly, we
can
keep on trucking and construct ever more elaborate incompressible
quantum
fluids in this way.

It may be verified that by this construction we obtain states outside the
hierarchy construction we will discuss
in the next section.  Thus, this construction is more general.

The most general $K$ matrix may be constructed by an iteration
procedure, which can be decomposed into two simple steps.\refto{wz2290}
In step A, two given matrices $K_1$ and $K_2$ are combined into
$K=\pmatrix{K_1&0\cr0&K_2\cr}$. In step B, we let $K\to K+C$, where
$C$ is
defined to be a matrix in which every entry is equal to 1. By combining
these two steps we can reach the most general $K$ matrix. In particular,
the
hierarchy construction is reached in a specific sequence of  step A's and
B's. See \ref{wz2290} for further details.

The $K$ matrix formalism provides a complete classification of abelian
quantum Hall fluids.\refto{wz2290} We mention in passing a classification
using the so-called $W$ algebra, the algebra associated with area
preserving diffeomorphisms.\refto{ctz} Since the Hall fluid is
incompressible, its area is by definition an invariant. Thus, perhaps to
nobody's surprise, the possible types of Hall fluids can be classified using
the $W$ algebra. It has been shown that this more mathematical
classification is in one-to-one correspondence with the classification using
the $K$ matrix.\refto{ctz}

Here I would like to describe another ``natural''
generalization.\refto{WZR, ETHE, wz2290}
We can
consider the possibility that each of the incompressible fluids
carries a charge $t_I$, not necessarily equal to 1.  Thus, in the effective
Lagrangian we will have
the couplings of the electromagnetic gauge potential $A_\mu$ to the gauge
potentials $\a_I$ described by
$$4\pi\cL=2A \sum_I t_I\eps \part \a_I +\ldots \eqno(12.30)$$

One motivation is simply ``why not?"  Another motivation is that the short
distance physics we have consistently ignored, and which we are unable to
treat, may produce bound states.  For instance, two electrons may bind into
a
charge 2 boson. The filling factor
$\nu={2\pi n\over eB}$ then scales to $\nu=\nu_{\rm old}(1/2) (1/2)
=\nu_{\rm
old}/4$.  Since the density is half of what it was and the charge (which we
have indicated explicitly) is doubled.  Thus, for instance, the $\nu=1/2$
state
is transformed into a $\nu=1/8$ state for which a Laughlin-type wave
function
may be written down readily
$$\psi\sim \prod_{i>j} (z_i-z_j)^8 e^{-\sum |z|^2} \eqno(12.31)$$
This was discussed by Halperin long ago and by Wen and Zee\refto{WZR}
 in the language of
effective Lagrangians more recently.
The effective Lagrangian is then
$$4\pi \cL =8\a \eps\part \a +2 A\ (2\eps\part\a) \eqno(12.32)$$
Note that we cannot simply scale the gauge potential $\a$ because of its
coupling to the vortices.

With \(12.30) we find, as generalizations of \(11.22), \(11.24), and
\(11.26),
 the filling factor and Hall
conductance to be
$$\nu=\si_H =\sum t_I K_{IJ}^{-1} t_J,\eqno(12.33)$$
the charge to be
$$q=\sum t_I K_{IJ}^{-1} l_J,\eqno(12.34)$$
and the statistics to be
$${\th\over\pi} = \sum l_I K_{IJ}^{-1} l_J.\eqno(12.35)$$
I find these formulas more symmetric and attractive than those in
\(11.22), \(11.24), and \(11.26) where we see by hindsight that we had set
the $t_I$'s to 1.  Here the $t_I$'s and the $l_I$'s
appear on equal footing.  The conductance has to do only with the $t_I$'s,
the
statistics only with the $l_I$'s, the charge with both.

In discussing the hierarchy states, it is often more convenient to go to a
basis in which $t_1 = 1$ and all the other $t_I$'s ($I>1$) equal to zero.
This simply corresponds to rotating to another basis. In this so-called
hierarchy basis the $K$ matrix has the form\refto{wz2290, bw}
$$
K^h=
\pmatrix{
p_1 & 1 & 0   & \cdots & 0\cr
1 & p_2 & 1   &        & \vdots\cr
0 & 1 & \ddots & \ddots & 0\cr
\vdots & &\ddots   & \ddots & 1\cr
0 &\cdots   & 0 & 1    & p\kappa
\cr}
\eqno(khiera)
$$

Let us now go back to \(12.34) and \(12.35) take $l_I$ to be the columns in
$K$, in other words, take
$$l_J^{(L)} =K_{JL} \eqno(12.36)$$
Then, we find from \(12.32) and \(12.35)
$$q^{(L)} = \sum_I t_I \de_{IL} =t_L\eqno(12.37)$$
and
$${\th^{(LL')}\over\pi} =\sum_{IJ} K_{IL'} K_{IJ}^{-1} K_{JL}
=K_{LL'}
\eqno(12.38)$$
These formulas give the physical meaning of $t$ and $K$.
We also now see clearly and immediately that there is a hole in the
excitation
spectrum if some $t_L$ is equal to 1 and if $K_{LL}$
(no sum on $L$) is an odd integer. We also
see that the statistical phase acquired by an ``electron'' of the $L$th type
moving around a vortex of the $L'$th type is given simply by
$${\th^{(LL')}\over\pi} =\de_{LL'} \eqno(12.39)$$
We merely computed $\th$ using \(12.35) and taking $l_J^{(L)} =
K_{JL}$
and
$l_J^{(L')}=\de_{JL}$.  We put quotation marks around the word electron
because this excitation {\it would} be an electron (or a hole) only
if $t_L=1$ and $K_{LL}$ is odd.

Now that we have learned that an ``electron'' of the $L$th type consists of a
bound state of $K_{JL}$ vortices of the $J$th type, we can go to a new
basis.
Write
$$j_{\rm vortex}^J =K_{JL}\ j_{\rm ``electron''}^L \eqno(12.40)$$
Then the coupling of the gauge potentials to the vortices may be written as
$$\a_J\ j_{\rm vortex}^J = \a_J\ K_{JL}\ j_{\rm ``electron"}^L
\eqno(12.41)$$
In other words, we should use the gauge potentials
$$\be_{\mu L} =\sum_J K_{JL} \a_{\mu J} \eqno(12.42)$$
In this basis, the Chern-Simons Lagrangian governing the gauge interaction
$$4\pi\cL=\a K\eps\part \a \eqno(12.43)$$
becomes
$$\eqalign{ 4\pi\cL &= (\be K^{-1}) K \eps\part (K^{-1}\be)\cr
&= \be K^{-1} \eps\part \be \cr}\eqno(12.44)$$
The matrix $K$ has exchanged places with its own inverse $K^{-1}$!  We
have
discovered the duality between the vortex basis and the ``electron'' basis.

\head {XI. Hierarchy}
\taghead{11.}

Our construction is general but abstract. The hierarchy states, as originally
constructed by Haldane, Halperin, and Laughlin,\refto{haldaneh, halperinh,
laughlinh}
 is based on the rather
physical
idea that the quasiparticles, being charged, could themselves condense and
produce an incompressible quantum state. Let us now try to implement this
construction using the formalism described here. Start with the Lagrangian
 \(cop) describing
the
$\nu=1/k$ state, with $k$ odd
$$
\cL ={1\over4\pi} (k\alpha\varepsilon\partial\alpha+2A\varepsilon\partial
\alpha) +
|(\partial-i\alpha)\phi|^2
\eqno(hhh)$$
The last item describes the coupling $\alpha j$ of the gauge potential
$\alpha$
to the vortex field $\phi$. (For simplicity  of notation, we write it in a
schematic relativistic form. The following discussion is not changed one bit
if
we were to write out the full non-relativistic form.)
 Note two points. (1) $\phi$ is a bose field; the
statistics of the vortex is induced by
 its coupling to $\alpha$. (2) $\phi$ does
not couple to $A$ directly; the charge of the vortex comes from its
coupling to
$\alpha$.

In general, the field $\phi$ may be bound to an even units of quantum flux
and still behaves as
 a bose field. This possibility may be implemented (see \(fx))
 by replacing the  last term in \(hhh) by
$$
|(\partial-i\alpha-i\beta)\phi|^2+{1\over4\pi p}
 \beta\varepsilon\partial\beta
\eqno(hbt)
$$
with $p$ an even integer.

Now we are ready to let $\phi$ condense. After it condenses, we dualize
the
phase degree of freedom according to the discussion in section VI. The
field
$\phi$ is
replaced by a gauge potential $\alpha'$ and the expression in \(hbt)
becomes
$$
{1\over4\pi}\Big[2(\alpha+\beta)\varepsilon\partial\alpha'+{1\over p}
\beta\varepsilon\partial\beta\Big]
\eqno(haa)
$$
After integrating over $\beta$, this becomes
$$
{1\over4\pi}\Big[2\alpha\varepsilon\partial\alpha'+p\alpha'\varepsilon
\partial\alpha'\Big]
\eqno(hbb)
$$
Adding this to  the first two terms in \(hhh) and combining $\alpha$ and
$\alpha'$ into a two-component gauge potential $a$
 we finally obtain a
Lagrangian of precisely the same form as in \(cop) as we must:
$$
\cL={1\over4\pi}(a K \varepsilon \partial a + At\epsilon\partial
a)
\eqno(hcc)
$$
with
$$
K= \pmatrix{k&1\cr 1&p\cr}
\eqno(hdd)
$$
and
$$
t=\pmatrix{1\cr0\cr}
\eqno(hee)
$$
This leads to $K^{-1}={1\over pk-1} \pmatrix{p&-1\cr -1&k\cr}$
and $\nu=tK^{-1}t={p\over pk-1}={1\over k-{1\over p}}$
with $p$ even. This is precisely the hierarchy state. For instance, with
$k=3$
and $p=2$  we obtain the $\nu=2/5$ state.

One interesting point is that we must attach flux to the vortex field $\phi$
before allowing it to condense. It not, $p=0$, and we do not get a Hall
state.
This is in fact a generic phenomenon worth commenting
 on in more detail. Consider
the
Lagrangian
$$
\cL=|(\partial-i\alpha)\phi|^2
\eqno(hff)
$$
(that is, we focus on the last term in \(hhh)).
 If $\phi$ condenses, the standard
Higgs mechanism operates and we obtain a massive gauge boson $\alpha$:
$\cL\to
v^2\alpha^2$
 as is familiar. Instead, suppose we attach an even units of flux
to $\phi$  so that we  have \(hbt). Letting $\phi$ condense and the Higgs
mechanism operate we obtain
$$
\eqalign{
\cL &= v^2(\alpha+\beta)^2+{1\over4\pi p} \beta\varepsilon\partial\beta\cr
&=v^2\alpha^2+2v^2\alpha\beta+\beta(v^2+{1\over4\pi p}
\varepsilon\partial)\beta}
\eqno(hii)
$$
Integrating out $\beta$ and using \(toy4) as always, we obtain
$$
\eqalign{
\cL &=v^2\alpha^2-v^4\alpha ({1\over v^2+{1\over4\pi
p}\varepsilon\partial})
\alpha\cr
&= {1\over4\pi p}\alpha\varepsilon\partial\alpha+\dots}
\eqno(hjj)
$$
where the dots indicate a long distance expansion. The would-be Higgs
mass
cancels out!
 (This is of course the same
as \(hbb) after we integrate out $\alpha'$).

Thus, in one case, we obtain the familiar Higgs mechanism  and a massive
gauge boson. In the other, we obtain a gauge potential $\alpha$ obeying
Chern-Simons dynamics. What is going on? Naively, when we think of
attaching
even units of flux to point  particles,
 we think of infinitely thin lines of flux, and
we suppose naively that we have not done anything substantive. As each
point
particle moves around another, it picks up a phase
$e^{i\pi p}=1$, that is, no phase at all. But in fact the subtle phase
relation contained in the wave function has been perturbed. This shows up
clearly when the field condenses: in effect, we smeared the thin lines of
flux
into a uniform background of flux. That the Lagrangian in \(hbt) is really
different from the Lagrangian in \(hff) is also clear by their symmetry
property.
In \(hff), we have a $U(1)$ invariance, while in \(hbt), we have a
$U(1)\times
U(1)$
invariance. Incidentally, these symmetry considerations essentially
determine
the form of the Lagrangian after condensation. Thus, in \(hjj) we should
have a
$U(1)$ invariance left and thus the would-be Higgs mass term
$v^2\alpha^2$
better cancels out.

This whole story is reminiscent of a discussion of the $O(3)$
nonlinear $\sigma$-model in $2+1$ dimensional spacetime given by Wen
and Zee.\refto{WEZ}
They argued that a Hopf term with an integer coefficient
 can always be added to the
Lagrangian: the effect in the path integral is to insert a factor formally
equal to 1 since suitable normalization is quantized in units of $2\pi$.
However, after a mean field approximation, one finds that the physics in
fact
depends on the coefficient of the Hopf term and that parity $P$ and time
reversal $T$ are broken just as in \(hjj).

\head {XII. Magnetic Monopoles and Multi-layered Hall Systems}
\taghead{12.}

Recently, experimenters are able to construct and study multi-layered
quantum Hall systems,\refto{B,T,E, murphy} a remarkable feat that has
generated
considerable interest in the theoretical community.\refto{YMG, MYG,
MPB,RM,FG, R} As
we will see presently, the
$K$ matrix formalism we
constructed here is ready made to study such systems. We will focus here
on the double-layered system, which for our purposes we can think of
simply as two parallel planes
in which electrons move in the presence of a magnetic field perpendicular
to the planes. As before, the electrons are supposed to be spin polarized so
that they are effectively spinless fermions. We will give a brief treatment
here, once again referring the reader to the original literature for
details.\refto{wzmono1, wzmono2, wzmono3}

Let us begin by identifying two relevant energy scales.  We can model the
two layers by a double well potential. By elementary quantum mechanics,
the energy gap $\Delta_{SAS}$ between the symmetric
and the anti-symmetric wave functions in the double wells  measures the
electron tunnelling amplitude between the two
layers. The other energy scale is determined by the interaction energy $V$
between electrons. For $\Delta_{SAS}$ large compared to $V$,
electrons are essentially tunnelling freely between the two layers, and the
double-layered system reduces to a single-layered system  Thus, we
will  study the opposite limit,
with $V \gg \Delta_{SAS}$, namely the weak tunnelling limit.

We proceed with general symmetry considerations. In the absence of
interlayer tunnelling, $N_1$ and $N_2$, the number of electrons in each
of the two layers, are separately conserved. Thus, the system has a
$U(1) \times U(1)$ symmetry, associated with the conservation of
$N_1+N_2$ and
$N_1-N_2$. The $U(1)_{N_1-N_2}$ symmetry may be spontaneously
broken due to quantum correlations between the two layers, in which case
we have a gapless Nambu-Goldstone mode. The symmetry may also be
explicitly broken due to electron tunnelling between the two layers, in
which case we expect the gapless mode to acquire a
gap. The analysis below confirms these expectations from general
symmetry considerations.

Our formalism involves a multitude of gauge potentials $a_I$. Recall that
they come in because their curls give conserved currents $J_I = {1\over
2\pi} \epsilon \part a_I$.  In the absence of tunnelling between the two
layers, the current associated with each layer, call it $J_I$ for $I=1,2$, is
separately conserved. The formalism is really ready made for studying
double-layered Hall systems! Thus, the simplest possible  double-layered
Hall
system is described by the effective Lagrangian
$$
\cL={1\over 4\pi}( \sum_{IJ} K_{IJ} a \eps \part a +\sum_I 2A\eps \part a
)+
a_1 j_1+a_2 j_2 +
\hbox{Maxwell  terms}
\eqno(5)$$
with a 2-by-2 matrix $K=\pmatrix{l&n\cr n&m\cr}$ The filling factor
$\nu=(l+m-2n)/(lm-n^2)$ is given by \(11.22). We note in
passing that this $K$ matrix is in one-to-one correspondences with a class
of wave functions proposed long ago by Halperin\refto{Hlp} to describe
double-layered Hall systems.

For a general $K$ matrix, we have an incompressible Hall fluid as before.
We can use the formulas derived in section IV to calculate the charge and
statistics of the quasiparticles. There doesn't seem to be much new here.

Let us now raise the interesting question of what would happen if $K$ has
a zero eigenvalue. As explained in section II, the coefficient of the
Chern-Simons term is proportional to the mass of the gauge field. Thus, in
this
case, some linear combination of
the
gauge fields becomes massless, with dynamics governed by Maxwell terms
in \(5).
In particular, for $K=k \pmatrix{1&1\cr 1&1\cr}$ (with $\nu=1/k$ as we
can see by taking a
suitable limit),
the combination $a_+=a_1+a_2$ has finite gap and couples
to the electromagnetic
potential $A_\mu$ thus describing a Hall fluid, while the
combination $a_-=a_1-a_2$ is gapless with a linear dispersion,
describing
a superfluid associated with fluctuations in $n_1-n_2$, the relative number
density between the two layers. Note that
$a_-$ decouples
from the electromagnetic potential and thus represents a neutral mode.

Our effective Lagrangian
\(5) becomes
$$
\eqalign{
\cL &= {1\over 4\pi} k(a_1+a_2)\epsilon\part (a_1+a_2) + {1\over 2\pi}
A\epsilon\part (a_1+a_2) + {1\over2}(a_1+a_2)(j_1+j_2)+\cr
&+{1\over2}(a_1-a_2)(j_1-j_2) + \hbox{Maxwell terms}
\cr}
\eqno(effl)
$$
Note that the third and fourth terms are just a trivial rewrite of the terms
$a_1 j_1+a_2 j_2 $ in \(5) above. The first three terms in \(effl) represent
nothing but our generic effective Lagrangian for the Hall fluid specialized
to a one-by-one $K$ matrix given by $k$. Thus, we see immediately that
the quasiparticles carry charges $\pm {1\over 2k}$. In particular, for
$k=1$ they carry half charges $\pm {1\over 2}$. The factor of $1\over 2$
comes from elementary arithmetic. This result was rederived
more recently using a more elaborate formalism.\refto{yang}

Physically, we can easily understand the appearance of the gapless mode
$a_-=a_1-a_2$ as due to the coherent fluctuation of  flux and
density.\refto{WZ2,Wil} Notice that the electrons in the second layer
behave
like flux tubes of $-2\pi$ flux to the electrons in the first layer.
 Recall that
$2\pi$ is the flux quantum in our units. (Let us treat $k=1$ for simplicity.)
Therefore the electrons in the first layer
see an effective magnetic field $B^*=B-2\pi n_2=(1-\nu_2)B$, where
$\nu_1$
and $\nu_2$
(with $\nu_1+\nu_2=1$)
 are the electron filling fractions in the two layers. The electrons in the
first layer think that they are in a state with  filling factor
$\nu^*_1={2\pi n_1\over B^*}={\nu_1\over 1-\nu_2}=1$ and are happy.
Similarly  $\nu^*_2=1$ and the electrons in the second layer are also
happy. Now consider a density wave such that the oscillations in the two
layers are of equal amplitude but out of phase. In other words, we consider
a wave such that whereever the density in the first layer has a crest, the
density in the second layer has a trough, and vice versa. Now note that as
$n_1$ goes up,
 $B^*$ also goes up since $n_2$ goes down. The effective filling factor
seen by the electrons in the first layer remains at  $\nu^*_1 =1$. Similarly
for the effective filling factor seen by the electrons in the second layer.
Thus, the electrons remain happy. In other words, as the wavelength of this
wave goes to infinity, the energy goes to zero. The mode is gapless. The
coherent fluctuation of flux and density is responsible for the gapless mode.
 In contrast, in the usual quantum Hall fluid,  the magnetic field
is fixed by the experimentalist, and
thus a density fluctuation
will change the filling factor and hence has a finite energy gap.
Here the effective magnetic field and density can fluctuate
together without changing the effective filling factor.
We note in passing that a similar mechanism is responsible for  anyon
superconductivity.\refto{WZ2}

Thus far, we have not included tunnelling. How is tunnelling represented in
this picture? When an
electron
tunnels from one layer to another, the currents $J_1^\mu = {1\over
2\pi}\epsilon\part a_1$ and $J_2^\mu= {1\over 2\pi}\epsilon\part a_2$ are
no longer separately conserved. Since electrons are represented by flux
quanta, tunnelling from the first layer to the second converts flux of type
$\epsilon\part a_1$ to flux of type  $\epsilon\part a_2$. Thus tunnelling
corresponds to a kind of magnetic monopole into which flux of type 1
disappears and out of which flux of type 2 appears. Indeed, more formally,
we have
$$
\Delta(N_1-N_2)=\pm 2 =
\int dt d^2x \part_\mu
(J_1^\mu -J_2^\mu)={1\over 2\pi}\int dt d^2x \part_\mu(\eps^{\mu\nu\la}
\part_\nu a_{-,\la})
\eqno(7)$$
Suppose we continue from Minkowskian (2+1) dimensional spacetime to
Enclidean $3$ space. We recognize $\part_\mu(\eps^{\mu\nu\la}
\part_\nu a_{-,\la})$ in \(7) as
$\vec\nabla\cdot(\vec\nabla\times\vec A)=\vec\nabla\cdot\vec B$
if we identify $a_{-,\lambda}$ as a 3-vector gauge potential $\vec A$ in
Euclidean space and $\vec B$ the corresponding magnetic field.
 This is precisely a Dirac
magnetic monopole, with its flux quantized according to \(7) to be $\pm
4\pi$, just as Dirac said it should be. Again, we mention in passing that in
discussions of anyon superconductivity magnetic  monopoles
convert anyon superfluid into normal fluid.\refto{WZ2}

Thus, in Euclidean $3$ space we have a plasma of magnetic monopoles and
anti-monopoles. At the location of each monopole and each anti-monopole
there occurs a tunnelling event in spacetime. Polyakov\refto{pkv}
showed long ago
 that in the presence of a monopole plasma the photon acquires a mass.
In the present context, this means the gauge potential $a_-$ acquires a
mass gap, in agreement with our expectations from general symmetry
considerations.

Armed with our effective Lagrangian, we can now go on to study various
phenomenological implications within the framework of linear response
theory. Again we refer the reader to the original
literature.\refto{wzmono1,wzmono2,wzmono3}

Let us sketch how Polyakov's analysis goes. The grand partition function of
the monopole plasma is given by
$$
Z\sim\sum^\infty_{N=0} {\zeta^N\over N!}\prod_a \Big(\int d^3x_a
\sum_{q_a=\pm1}\Big) e^{-\sum_{bc}{1\over g^2}{q_bq_c\over|x_b-
x_c|}}
\eqno(gp)
$$
Reading from right to left, we see  the Boltzmann factor, given by the
exponential of the Coulomb interaction, expressed in terms of the magnetic
charge $q_a = \pm 1$ and position $x_a$ of the $a$-th monopole, $a=1, 2,
... N$. We next sum over the charges and positions. The entire expression
is then weighted by a fugacity factor. The fugacity $\zeta$ tells us the
probability of having a monopole in a unit volume of spacetime and is thus
proportional to the probability of a tunnelling event per unit space and per
unit time. Introducing the density of monopoles
$$
\rho(x) = \sum_a q_a \delta^{(3)} (x - x_a)
\eqno(density)
$$
we can rewrite the Boltzmann factor in \(gp) as a functional integral
$$
\int \cD\theta e^{i\int d^3x[{1\over2}g^2(\partial\theta)^2+\theta\rho]}
\eqno(sine)
$$
We easily verify this representation by integrating over $\theta$ and using
\(toy4) and \(density). Summing over the charge and position of the
monopole we obtain
$$
\int d^3 x_a\sum_{q_a=\pm1} e^{iq_a\theta(x_a)}=2\int d^3x\cos\theta(x)
\eqno(cos)
$$
Finally, we sum over the number of monopoles
$$
\sum^\infty_{N=0} {\zeta^N\over N!} [\int d^3x\cos\theta(x)]^N=
e^{\int d^3x\zeta\cos\theta}
\eqno(lift)
$$
Thus, we obtain
$$
Z\sim\int \cD\theta e^{i\int d^3x[g^2(\partial\theta)^2+\zeta\cos\theta]}
\eqno(partti)
$$
The effective Lagrangian is sine-Gordon
$$
\cL_{eff}=g^2(\partial\theta)^2+\zeta\cos\theta
\eqno(sg)
$$

Expanding the cosine term in \(sg) we see that indeed the $\theta$ field has
mass $\sim \sqrt{\zeta}/g$.

We make the important remark that the field $\theta$, that is, the ``order
parameter", is an angular variable. Looking back to \(cos) we see that the
angular character of $\theta$ comes about because the magnetic charge of
the monopole is quantized.

It is immensely pleasing that some of the most profound concepts in
theoretical physics are involved here. The discretness of the electron leads
to Dirac quantization of the magnetic monopoles (\(7)). The quantization of
monopoles leads to an angular variable as the order parameter.

The appearance of an angular order parameter immediately reminds us of
the Josephson effect in superconductors. Wen and I\refto{wzmono1}
 were thus led to predict
that there should be ``Josephson-like" effects in double-layered quantum
Hall systems. We were careful to use the term ``Josephson-like" because the
double-layered Hall system is to be sure not a superconductor, and thus the
usual discussion of the Josephson effect must be taken over here with great
care. A detailed discussion is beyond the scope of these lectures. We refer
to the original work\refto{wzmono1, wzmono2, wzmono3} and to the
recent  literature.\refto{yang,moon,ich}

Let us remark in passing that an interesting probe is provided by applying
a magnetic field parallet to the plane of the double-layered Hall system.
When an electron tunnels from one layer to another, its wave function
now acquires a phase factor
$$
e^{\pm ie\int dz A_z} \equiv e^{\pm i\xi(x)}
\eqno(tunphase)
$$
(We denote the coordinate perpendicular to the plane by $z$ and the
two-dimensional coordinates in the plane by $x$.) Think about the current
coil
generating the in plane magnetic field and you see easily that parity,
 that is, reflection in a mirror perpendicular to
the $z$ axis, is broken, thus accounting for the $\pm$ sign in \(tunphase)
according to whether the electron tunnels from the first layer to the
second or vice versa. Thus, in \(gp), a monopole is now associated with the
phase factor $e^{+i\xi(x_a)}$, and an anti-monopole with the phase factor
$e^{-i\xi(x_a)}$. Tracking through the analysis given above, we see that
in \(cos) we would pick up this additional phase factor with the
consequence that in \(sg) the $cos \theta$ term  is now replaced by $cos
(\theta + \xi)$. Wen
and Zee\refto{wzmono2} considered a random magnetic field and showed
that its effect was to reduce the tunnelling parameter $\zeta$. Yang et
al\refto{yang} showed that a uniform magnetic field would drive an
interesting commensurate-incommensurate transition.

\head {XIII. Shift and Spin}
\taghead{13.}

One topic that we will mention here only in passing is that of the
topological quantity called the shift and its relation to orbital spin. The
discussion will be so brief as to be merely a sketch. The reader is referred
to the original literature for details.\refto{shift} Let me
motivate the discussion first physically and then mathematically.

Physically, we know that electrons in the second Landau level have more
energies than electrons in the first Landau level. In the semi-classical
treatment in Section~I,
 we would endow the electrons in the second level with one
more quantum of angular momentum than those in the first level. We may
picture  electrons in different Landau levels as moving around in Larmor
orbits of different sizes. In our effective Lagrangian treatment, however,
we simply represented the degree in the $I$-th Landau level by a scalar
field $\phi_I$ (recall \(11.12)) which later metamorphosed into a gauge
potential \(11.13).
 We would like our theory to be smart enough  to recognize the
different angular momentum properties associated with different gauge
potentials $\a_I$.

A clue on how to incorporate the angular momentum properties comes
from studying non-interacting (spinless)
 electrons on a sphere. Put a magnetic
monopole of strength $G$ (which according to Dirac can be only a half-
integer or an integer) at the center of the sphere. The flux through the
sphere is equal to $N_\phi = 2G$. I leave it to you to work out the
elementary exercise\refto{haldane} of showing that the single electron
energy is given by
$E=({1\over2}\hbar
\om_c)\left[l(l+1)-G^2\right]/G$ with the Landau levels corresponding to
$l=G,G+1,G+2,\dots$. The degeneracy of the $l$th level is $(2l+1)$. Thus,
if
$L$ Landau levels are filled with non-interacting electrons we have
$N_e=\sum_{k=0}^{L-1}\left[2(G+k)+1\right]=LN_\phi+L^2$. Thus, we
see that with the filling factor $\nu=L$ the relation between $N_e$ and
$N_\phi$ is not given simply by  $N_\phi =\nu^{-1} N_e $ but by the
slightly more complicated relation
$$N_\phi =\nu^{-1} N_e - \cS \eqno(2hd)$$
where we have defined the topological quantity $\cS$ called the shift. For a
sphere, $\cS = \nu$.

Mathematically, we see that in our effective Lagrangian \(12.30) we couple
the
electromagnetic gauge potential $A_\mu$ to  the conserved current
$\sum_I t_I\epsilon \part \alpha_I$. On a
curved surface, there exists another ``gauge potential" (or, for those of you
familiar with the mathematical terminology of differential forms, a gauge
1-form), namely the connection $\omega_\mu$ whose curl gives the
curvature of the surface. Thus, in our effective Lagrangian  we can add
the term
$$
\Delta\cL={1\over 2\pi} \omega J_s
\eqno(spin)
$$
where we define the conserved current $J_s \equiv \sum_I s_I\epsilon
\part
\alpha_I$. In any case, we are always free to add such a term.

To be sure, real samples are flat and $\omega$ vanishes. Nevertheless,
$\omega$ gives us a valuable probe into the Hall fluid. We know from
basic physics that a spinning particle moving along a loop in curved space
acquires in its wave function a phase factor
$$
e^{is \oint \omega}
\eqno(phs)
$$
just as a charged particle moving in a magnetic field acquires in its wave
function an Aharonov-Bohm phase factor
$$
e^{it \oint A}
\eqno(pht)
$$
Here $s$ and $t$ measures the effective spin and charge of the particle.
(We should emphasize that in the treatment of the Hall fluid given here, we
are always dealing with spin up electrons or effectively spinless electrons,
as we mentioned in section I. The spin we are talking about here is the
orbital spin of the electron.) Just as the variation of the effective
Lagrangian with respect to $A$ tells us about the electromagnetic
properties of the Hall fluid (for example, the charges of the quasiparticles)
as we have discussed in detail, the variation of the effective Lagrangian
with respect to $\omega$ tells us about the angular momentum properties
of Hall fluid.

Proceeding in this way, we find a remarkable simple formula for the shift
$$\cS=2(1-g)\nu^{-1} (tK^{-1}s)\eqno(8shift)$$
The shift depends on the genus $g$, but not the metric, of the manifold,
and is
thus a topological quantity. For the sake of completeness, let me record
here that for certain specific Hall states the shift and the spin vector $s_I$
may be worked out explicitly.\refto{wz2290} On the sphere, for the state
described by $K$ in the symmetric basis (with $t_I = 1$, all $I$) the shift
is
given by
$$\cS={1\over \nu} \sum_{IJ}(K^{-1})_{IJ}K_{JJ}\eqno(10shift)$$ and
the  spin vector is given by
$$
s_I={1\over2}K_{II}
\eqno(11shift)
$$
For the hierarchical states described by $K^{h}$
in the hierarchical basis (with $t_I=\delta_{I1}$) we have
$$ \cS={1\over \nu}\sum_I[(K^{h})^{-1})]_{1I}K^h_{II}\eqno(12shift)$$
and thus
$$s_I={1\over2}K^h_{I1}\eqno(13shift)$$
For instance, for the hierarchical state characterized by
$K^h=\pmatrix{p_1&l\cr l&p_2\cr}$ and filling factor $\nu=(p_2-
l^2/p_1)^{-1}$
we have $\cS=p_1-l$.
(The $\nu=2/5$ state is described by $K^h=\pmatrix{3&-1\cr-1&2\cr}$
and $\cS=4$.
The $\nu=2/7$ state is described by $K^h=\pmatrix{3&1\cr1&-2\cr}$ and
$\cS=2$.)
 From the above we see that the spin vector in the
(generalized) hierarchical FQH states may be quite complicated.
However, for the
Laughlin states with filling fraction $1/m$ we have the simple result
$s=m/2$ and $\cS=m$.

These results proved to be useful in distinguishing between different Hall
fluids with the same filling factor but different internal angular momentum
properties.\refto{dhwen}

\head{XIV. The Even Integers}
\taghead{14.}

Starting this last lecture, I feel a bit frustrated as there are
still a lot of
material on the quantum Hall fluid to be covered. Instead of rushing
through a list of topics, I have decided to start a more or less
completely  new subject. So the student who has not understood the
previous seven lectures can relax and enjoy a fresh start.

The title of  today's lecture is  ``Does Mother Nature know about the even
integers?"  Less flippantly and  more precisely, we have seen that the
topological  fluids described  by
$$
\cL={k\over4\pi} a\varepsilon\partial a . \eqno(inte)
$$
for $k=1, 3, 5...$ are realized in the Hall
system with filling factor $\nu=1,1/3,1/5$, \dots Naturally, we are tempted
to
ask whether  the topological fluids for
$k= 2, 4, 6,...$ are also realized in
Nature?

Let us start with chapter 1 of any solid state physics texts or
Feynman's
freshman lectures. Consider a spinless particle hopping on a
square two
dimensional lattice:
$$H=-\sum \left\{ c_{i+x}^\dagger c_i
+c_{i+y}^\dagger
c_i+h.c.\right\}\eqno(3.8)$$
(We write $i+x$ for
$(i_x+1,i_y)$\etc) We denote the
coordinates on the lattice by
$i=(i_x,i_y)$.  Our
lattice spacing is fixed to be 1 and $i_x, i_y$ are
integers.
We suppress the
two
dimensional character of $i,j,$, \etc We also trust the reader to
distinguish
between the site location $i$ and the imaginary unit $i$.
We Fourier-transform to momentum space
$c_i={1\over\sqrt
N} \sum_k c_k e^{-iki}$ where as usual $k={2\pi\over
N}l$
with
$l=(l_x,l_y)$ with $l_x,l_y$ integers and $k$ ranges over the
Brillouin
zone $-\pi\le k_x\le \pi$, $-\pi\le k_y\le \pi$.    The
hopping
from $i$ to $i+\mu$ ($\mu=x$ or $y$) in real space
is
associated with $e^{ik_\mu}$ in momentum space because of the phase
mismatch
between $c_i$ and $c_{i+\mu}^\dagger$. We thus obtain
immediately
$$
H=-\sum \left\{ c_k^\dagger c_k e^{ik_x} +c_k^\dagger
c_k
e^{ik_y}+h.c.\right\}.\eqno(3.10)
$$
from which we can read off  the energy eigenvalues
$$
E(k) = -(\cos k_x + \cos k_y)
\eqno(eigen)
$$

Let us now
regard the particle as a fermion and fill the energy eigenstates
with non-interacting fermions. At half-filling, by symmetry, the states up
to $E=0$
are filled. The Fermi surface is defined by $E(k) = 0$ and is in fact
a
square  bounded by lines defined by $\pm k_x\pm
k_y=\pm\pi$
(uncorrelated sign). The reader should pause and draw the Brillouin zone
and
the Fermi surface.
 This is a typical
situation we
encounter in solid state physics.

We now follow Affleck and Marston,\refto{AM} Kotliar,\refto{K}
 and Ioffe and
Larkin\refto{IL} and suppose that there is a flux of $\Phi =
\pi$ threading
through each plaquette of the lattice. In other words,
suppose that every
time the particle goes around a plaquette, its wave
function acquires a
phase factor of $e^{i\Phi}=(-1)$. This seems innocuous
enough, but in fact, as
we will see, the phase  produces a drastic and
profound change in the
energy spectrum.   The Hamiltonian  is given by
$$
H=-\sum
\left\{ c_{i+x}^\dagger c_i(-)^{i_y} +c_{i+y}^\dagger
c_i+h.c.\right\}
\eqno(flux)
$$
Note the factor $(-)^{i_y} $.
This is the only difference between the
Hamiltonian in \(flux)
and in \(3.8). You should check that indeed whenever a particle goes
around a
plaquette its amplitude acquires a phase factor of $(-1)$.
(Some of you would recognize that this is a gauge theory written in a
specific
gauge.
This simply
means that \(flux) has the form
$$
H=-\sum_i \{c^\dagger_{i+x}U_{ix}c_i+c^\dagger_{i+y}
U_{iy}c_i+h.c.\}
\eqno(gga)
$$
and  that we are free to make the substitution $c_i
\rightarrow e^{i
\theta_i} c_i$, $U_{ix}\to e^{i\theta_{i+x}}U_{ix}e^{-i\theta_i}$, and
$U_{iy}\to e^{i\theta_{i+y}} U_{iy}e^{-i\theta_i}$
 in \(gga)). Fourier transforming
as before we obtain immediately
$$H=-\left\{ \sum c_{k+w}^\dagger c_k
e^{ik_x} +c_k^\dagger c_k
e^{ik_y}
+h.c.\right\}\eqno(3.9)$$
where we have defined $w=(0,\pi)$.
In contrast to \(3.10), here the
Hamiltonian is not translation invariant: the
position dependence transfers
momentum.  With $(-)^{i_y}=e^{i\pi i_y}
= e^{iwi}$, we see that the flux
kicks in a momentum $w$.

To  diagonalize this Hamiltonian
in \(3.9),
we shift the dummy summation
variable by $k\to k+w$, thus obtaining
$$H=-\left\{ \sum c_k^\dagger
c_{k+w} e^{ik_x} -c_{k+w}^\dagger
c_{k+w}
e^{ik_y}
+h.c.\right\}\eqno(3.12)$$
Note the minus sign. The Hamiltonians in \(3.9)
and \(3.12)
 are identical.  Let us add
them and divide by 2 to
obtain
$$H=-\sum{}' (c_k^\dagger c_{k+w}^\dagger)
\cH\left(\matrix{c_k\cr
c_{k+w}\cr}\right)\eqno(3.13)$$
where
$$
\cH=\left(\matrix{ \cos k_y & \cos k_x\cr
\cos k_x& -\cos k_y\cr}\right).
\eqno(3.14)
$$
In \(3.13), the summation over $k$ should be
restricted to half of the
Brillouin
zone if one wants to be pedantic.  The
prime on the summation symbol
serves as
a reminder.
  Equivalently, we
can recognize \(3.13) for a given  $k$
as defining  a two-site hopping
problem in momentum space: a
particle hops from $k$ to $k+w$ and with
another hop gets back to
$k+2w=k$.

  The
eigenvalues are
$$E=\pm
\sqrt{\cos^2k_x+\cos^2k_y}.\eqno(3.15)$$
The spectrum has been
drastically altered from \(eigen).
 Here $E(h)=0$ only at isolated
points,
namely $k=k_*=\left({\pi\over 2},{\pi\over 2}\right)$ and
$k=k'_*=
\left(-{\pi\over 2},{\pi\over 2}\right)$ and equivalent points.
The
lines of zero $\pm k_x\pm k_y=\pm\pi$ defining the Fermi's surface
 has been squeezed into isolated zeroes, which I
will
call Dirac zeroes for reasons that will become clear presently.

Let us
expand $\cH$ around one of the Dirac zeroes, say $(\pi/2,\pi/2)$.
For
$k=k_*+q$, $q$ small, we have
$$\cH=-q_y\tau_3 -q_x\tau_1
\eqno(3.16)$$
Multiplying by $\tau_2$, we see that the energy and
momentum are related
by
$$\ga_0\cH-q_x\ga_x-
q_y\ga_y=0\eqno(3.17)$$
where we have defined
$$\ga_0=\tau_2,\quad
\ga_x=i\tau_3, \quad \ga_y=-i\tau_1.\eqno(3.18)$$
We recognize this as
the Dirac equation for a massless particle
$$\ga_\mu q^{\mu} \psi =0
\eqno(Dirac)$$ Remarkably, in a system that is
not even rotationally invariant, we find that
the $E\sim 0$ excitations satisfy the massless Dirac equation!

The electrons are happy. They are no longer some solid state electron
living
on an awkward looking Fermi surface: they are now fancy electrons
answering to Dirac.

But perhaps there exists a theorem that nobody is ever
completely happy. The electrons want to be happier: they want to be
massive Dirac electrons.

Before we figure out how we can make the electrons happier, let us
contemplate the massive Dirac equation
$$(\ga_\mu q^{\mu} - m) \psi =0 \eqno(massiveDirac)$$ in $(2+1)-
$dimensional spacetime. According to Einstein, for a massive particle we
can always go to its rest frame in which $q^0=m$ and $q^i=0$ so that the
Dirac equation becomes
$$m(\ga_0 -1)\psi =0 \eqno(proj)$$
This equation simply tells us that certain components of $\psi$ do not exist.
In the familiar case of $(3+1)-$dimensional spacetime, $\ga_0$ is a 4-by-4
 matrix and the Dirac equation projects out two components
of the four-component $\psi$, leaving us with two spin components.
Here in (2+1)-dimensional spacetime, however, $\ga_0$ is 2-by-2 and thus
one
component of the two-component $\psi$ is projected out, leaving us with
only one spin component.

What does it mean for a spin $1/2$ particle to have only one spin
component? It only spins one way and not the other. Parity and time
reversal are broken!

We see that from this simple symmetry consideration if we want to turn the
elementary excitations into something that obeys the massive Dirac
equation we must break parity and time reversal. But, you may ask, by
introducing flux through the lattice, didn't we already break parity and
time reversal? The answer is no, because of the deep mathematical identity
$$
e^{i\pi}=e^{-i\pi} !
\eqno(pi)
$$
In other words, when our particle goes around the square clockwise, its
wave function picks up a phase of $(-1)$, but when it goes around the
square anti-clockwise, its wave function picks up the same phase.

But we know
$$
e^{i\pi\over 2}=e^{-{i\pi\over 2}}
\eqno(pi/2)
.$$
Thus, if we introduce a flux of $\pi/2$ in some suitable fashion, we would
break parity and time reversal. This crucial step was taken by Wen,
Wilczek, and Zee\refto{WWZ}, who added to the Hamiltonian \(3.8) a
term
$$\D H=\be \sum_i c_{i+x+y}^\dagger c_i  (+i) (-1)^{i_y}
+h.c.\eqno(3.19)$$
where $\be$ is a real constant. It is easy to check that a particle going
around half of a plaquette, namely a triangle (travelling clockwise $i\to
i+x+y\to i+x$) acquires a quantum phase of
be $e^{i\pi/2}$. Having gone through the discussion given earlier we can
write down
$\D H$ in momentum space immediately:
$$\D H=\be \sum c_{k+w}^\dagger c_k (+i) e^{i(k_x+k_y)} +h.c.
\eqno(3.21)$$
After shifting, we see that we obtain a term
$$\be c_{k+w}^\dagger c_k (+i) e^{-i(k_x+k_y)}\eqno(3.21')$$
to go with the term in \(3.21).  Under hermitean conjugation the $i$ flips
sign
while the shift of $k$ by $w$ produces another sign, so that we again
obtain a
cosine upon adding the two terms.  Thus, with the same basis as in \(3.13)
we
have
$$
\D\cH=\left(\matrix{ 0 & -i\be \cos (k_x+k_y)\cr
i\be\cos (k_x+k_y) &0\cr}\right)
\eqno(3.22)
$$
(Again, we ignore irrelevant overall real factors such as factors of 2.)  The
fact that we hop to our next nearest neighbor rather than the nearest
neighbor
produces $\cos(k_x+k_y)$ rather than $\cos k_x$ or $\cos k_y$.  This is
crucial, because then $\D\cH$ is just a constant near a Dirac zero, say
$\left({\pi\over 2},{\pi\over 2}\right)$, so that we now have, instead of
\(3.16)
$$\cH= -q_y\tau_3 -q_x\tau_1 -m\tau_2\eqno(3.23)$$
where $m$ is $2\be$ or something like that.

We obtain a massive Dirac fermion!  In condensed matter language, a gap
opens
in the energy spectrum.

Some of you might remember the Peierls instability
discussed in Wiegmann's lectures in this volume. The dynamical opening of
a gap there in a (1+1) dimensional context is closely related to the
opening of a gap here in a (2+1) dimensional context.

The dynamics of the low energy excitations are thus described by
$$
\cL = \sum\bar\psi (i\slp -m)\psi
\eqno(massive)
$$
The sum reminds us that there are actually two modes, one at
$\left({\pi\over 2},{\pi\over 2}\right)$ and one at $\left({\pi\over 2},-{
\pi\over 2}\right)$.

Let us now study phase fluctuations around this state. In the Hamiltonian
we replace $c^{\dagger}_i c_j$ by $c^{\dagger}_i c_j e^{i a_{ij}}$ and
treat the phase field $a_{ij}$ which lives on the links of the lattice as a
dynamical field.  Under a gauge transformation $c_i\to e^{i\th_i} c_i$, the
Hamiltonian is now unchanged if we take $a_{ij}\to a_{ij} +\th_i-\th_j$.
By
gauge invariance, we can argue immediately that in the long distance limit
the
effective Lagrangian \(massive) is now     replaced by
$$\cL = \sum\bar\psi (i\slp -m-\nna)\psi\eqno(4.1)$$
where $a_\mu(x)$ is a gauge potential. (The phase fluctuations $a_{ij}$ go
into
the spatial components of $a_\mu(x)$ as usual.  The time component
$a_0(x)$
emerges in a treatment more careful than the one given here.  It originates
in
a Lagrange multiplier term $\sum_{i} a_{0i} (c_i^\dagger c_i-1)$ we have
to add
in order to enforce the constraint of one particle per site when we fill the
single particle states with non-interacting fermions.)

Since the fermions are massive, at energies low compared to $m$, we can
integrate out the fermions to obtain an effective theory expressed solely in
terms of the gauge potential $a_\mu$. We have to calculate a simple one
loop
Feynman diagram, the analog of the so-called
vacuum polarization graph in quantum
electrodynamics.  The relevant integral is proportional to
$$\int d^3p\ tr\ \left[\ga_\mu {1\over\lnp+\lnq-m} \ga_\nu
{1\over\lnp-m}\right]
 \eqno(4.3a)$$
We are only interested in the low energy long distance physics and thus we
may
expand in powers of the  momentum $q_\lambda$.
  Set $q_\la$ to zero after differentiating
with
respect to $q_\la$ to obtain
$$\int {d^3p\over (p^2-m^2)^3}\ tr\ \left[\ga_\mu (\lnp+m) \ga_\la
(\lnp+m)
\ga_\nu(\lnp+m)\right]\eqno(4.3b)$$
This quantity is the coefficient of a term in the effective Lagrangian
quadratic in the gauge potential $a$ and linear in derivative. Do we know
such
a term?  Indeed it is none other than our old friend the Chern-Simons
term!

I will leave it to the reader to evaluate this integral in full.  Let us simply
look at one piece of the integral, say the piece coming from the term in the
trace proportional to $m^3$, thus
$$m^3\int {d^3p\over (p^2-m^2)^3} \eps_{\munu\la} \eqno(4.4)$$
First, we see that the antisymmetric symbol appears
as it must: in (2+1)-dimensional
spacetime $tr\ga_\mu\ga_\nu\ga_\la$ is proportional to
$\eps_{\munu\la}$.
Thus, in the effective Lagrangian we do obtain the Chern-Simons term
$\eps^{\munu\la} a_\mu \part_\nu a_\la$ as expected.
  Incidentally, the appearance of
the
$\eps$ symbol indicates that $T$ and $P$ are manifestly violated.  Second,
by
dimensional analysis, we see that the integral in \(4.4) is up to a numerical
constant equal to $1/m^3$.  But be careful! The integral depends only on
$m^2$
and doesn't know about the sign of $m$.  The correct answer is
proportional to
$1/|m|^3$,
not $1/m^3$.  Thus, the coefficient of the Chern-Simons term is equal to
$m^3/|m|^3=m/|m|=$ sign of $m$, up to a numerical constant.

Now I have to tell you what the summation symbol in \(massive) and \(4.1)
means.
Thus far, we have treated the hopping particle as a non-interacting
fermion. We
will now be more specific and say that the fermion is in fact an electron
with
spin either up or down. So we
 actually have to sum over four species of
Dirac
fermions:  two Dirac zeroes in the Brillouin zone and the up and
down
spin electrons .

At this point we should be worried.  The coefficient of the Chern-Simons
term
is proportional to the sign of $m$ and thus the contributions of the four
different fermion fields to the Chern-Simons term may cancel.  Indeed,
recall
that the mass term comes from a term like $\cos(k_x+k_y)$, which takes
on
opposite signs at the two Dirac zeroes $k_*=(\pi/2,\pi/2)$ and
$(-\pi/2,\pi/2)$.  It looks like we are not going to get a Chern-Simons term.
But wait, we may still get it!  Recall that $\cH$ contains a term like
$(\cos
k_x)\tau_1$.  Thus, at one Dirac zero, $\tau_1$ is used to construct the
gamma
matrices, while at the other, $(-\tau_1)$ is used.  In other words, the
representations of the gamma matrices at the two Dirac zeroes are
conjugate of
each other.  The contributions of the four Dirac fields thus add and we
obtain
for the effective Lagrangian
$$\cL=4 \left(1\over8\pi\right)\veps^{\mu\nu\la}a_\mu\part_\nu
a_\la\eqno(4.5)$$
The topological fluid described by \(4.5) is known as a chiral spin
fluid.\refto{WWZ} Note the factor of four.

By now you should certainly know
 that particles coupled to a gauge potential with
Chern-Simons dynamics acquire fractional statistics.
Referring to \(4.7)  we see that
 the induced Chern-Simons term in \(4.5)  endows the excitations
in the
system with semion statistics $(\th=\pi/2)$.  The word semion, half-Latin
and half-Greek in origin, is perhaps particularly appropriate as the name
for a
particle whose statistics lies half way between fermi and bose statistics.
Note that careful accounting practices, according to which we are required
to sum over four fermion
fields in \(4.1), are crucial in obtaining semion statistics.

Ishikawa and his collaborators\refto{KI,ISKY} have given an elegant
representation
of the Feynman diagrams contributing to the coefficient of the
Chern-Simons term.
(Their argument was actually given in the context of developing a
rigorous theory of the integer quantum Hall effect. We will adapt and
simplify
their arguments for our rather limited purposes here.) The antisymmetric
part
of the Feynman integral in \(4.3b) may be written as
$$\int d^3p\ \veps^{\mu\nu\la} tr\left[{\partial S^{-1}\over \partial
p_\mu} S
{\partial S^{-1}\over\partial p_\nu}S{\partial S^{-1}\over \partial p_\la}
S\right]\eqno(4.8)$$
where the inverse propagator
$$S^{-1}(p)=\lnp+m\eqno(4.9)$$
and
$${\partial S^{-1}\over\partial p_\mu}=\gamma_\mu\eqno(4.10)$$
The mathematically well-schooled reader would immediately recognize this
integral as some sort of homotopic invariant. Indeed, using the Lagrangian
of
the differential form and
introducing the
standard
Cartan-Maurer form
$$dS^{-1}S\equiv {\partial S^{-1}\over\partial p_\mu}S
dp^\mu\eqno(4.11)$$
we have the integral
$$\int tr(dS^{-1}S)^3\eqno(4.12)$$
The Feynman integral in \(4.3b) is no garden-variety a-dime-a-dozen
integral: it is a very special integral.

Before we conclude that we have a homotopic invariant, we have to deal
with
several mathematical details. In the context of the original problem, the
integration region over $d^2p$ is a Brillouin zone and thus a torus in
mathematics. The integration over $p_0$ is over the band. Thus, we are
integrating over some sort of ``pinched torus" rather than the three-sphere
$S^3$.
Ishikawa \etal\  had to invoke some mathematical theorem asserting that it
is
homotopically equivalent to integrate over $S^3$ (assuming we have
already
performed some sort of Euclidean continuation). Furthermore, in
Euclidean space
we recognize that
$$S^{-1}(p)=ip_a\tau_a+m\eqno(4.13)$$
(where we sum over the three Pauli matrices) is up to an irrelevant overall
factor an element of $SU(2)$ (since $(SS^\dagger)^{-1}=p^2+m^2$). For
further
details and a more general treatment than given here, see.\Ref{KI}

Let me mention an immediate generalization:
we can consider the state with flux per plaquette equal to $\Phi=2\pi(p/q)$.
This may be achieved with $U_{iy}=1$ and $U_{ix}=e^{i2\pi(p/q)i_y}$.
(The preceding
 discussion  corresponds to $q=2$.) Going through
the
same steps as before, we now obtain a Hamiltonian of exactly the same
form as
the Hamiltonian in \(3.9) but with the crucial exception that the vector
 $w$ is
now equal to $(0,2\pi p/q)$, the momentum transferred by $U_{ix}$.
 We thus have a
$q$-site hopping problem in momentum space. The corresponding $\cH$
 (see \(3.14)) is
now a $q$ by $q$ matrix, thus giving us $q$ energy bands. As before, the
crucial question is how many Dirac zeroes there are at $E=0$. It turns out
that
there are $q$ Dirac zeroes for $q$ even, but none for $q$ odd. Actually,
this
result is guaranteed by an index theorem, as was discussed by
Kohmoto\refto{KMT} and by Wen and Zee.\refto{WZIX}

The consequence of having $q$ Dirac zeroes rather than $2$ is easy to
work out:
the coefficient of the Chern-Simons term in \(4.5) is scaled up by $q/2$ and
thus
the statistics angle of the excitations around this generalized flux state is
scaled down by $q/2$ so that
$$\theta=(\pi/2)(2/q)=\pi/q\eqno(4.19)$$

We are now ready to turn to the physics of  a fluid of semions.
First, notice that two semions do not make a fermion.
You may think that semions are half way
between bosons and fermions and so if I put two semions together they
might
make a fermion.  But in fact, suppose we move a bound state of
two semions half way around another
bound state of two semions,
thus  interchanging  the two bound states. Then we get a
phase.
Look at one bound state and focus on one of the semions in it. The semion
gets
a phase of $i$ going around each of the two semions in the other bound
state.
Now we are ready to calculate:
$2\times 2=4$.  We get the phase four times. Each time a semion goes
around
another semion we get a phase of $i$.  Since we get this phase 4 times we
get
$i^4$, which is equal to 1.  Thus, two semions actually
make a boson. (The attentive reader would remember that this is closely
related
to the physics behind \(11.26) and \(12.11).)

We may now indulge in  some handwaving arguments.  We all know that
fermions like to
stay apart and bosons like to stick together.  Semions, being half way
between
are more likely than fermions to pair.  When they do pair, they form
bosons,
whose condensation can then lead to superfluidity and superconductivity.
Just
think, if you were a fermion and you want to condense into the ground
state,
what could you do?  You have basically two strategies: (1) Find another
fermion
and pair with him or her, or (2) turn yourself into a boson.  Strategy (2) is
an attractive possibility, but in the systems under discussion you can
apparently make it only half way and turn yourself into a semion.

Much of the physics of anyons is contained in the following deep
mathematical
identity $\th=0+\th =\pi-(\pi-\th)$.  What is the mathematics trying to tell
us?  The first equality says that an anyon with statistics $\th$ is effectively
a boson (statistics 0) with a gauge interaction of ``strength'' $\th$; the
second equality says that it is also a fermion (statistics $\pi$) with a gauge
interaction of ``strength'' $-(\pi-\th)$.  The physics is made completely
clear
by looking at the 2-anyon problem.  In the center of mass, the
Schr\"odinger
equation has a centrifugal potential like $(\ell +\th/\pi)^2/r^2$ with $\ell$
an integer.  The smallest possible value of this potential is thus $0/r^2$ for
bosons, $1/r^2$ for fermions, and $1/4r^2$ for semions.  Anyons are like
bosons
with ``centrifugal repulsion'' between them or fermions with a ``centripetal
attraction''.  Thus, theorists have a choice.  They can begin either with a
fermi gas\refto{TOS1,TOS2,TOS3}
 --- certainly not a superfluid ---  and show that the attraction
gives pairing and superfluidity, or with a boson
gas,\refto{TOS4,TOS5,TRUG}
 --- which with a short
ranged repulsion is a superfluid as was shown by Bogoliubov\refto {Bogo}
ages ago --- and
show that the repulsion does not destroy superfluidity.

Further discussions of the semion superfluid would take us far from the
announced subject of these lectures. The story of the semion superfluid is
another story for another day, and a darn fascinating story too, I may add.
I refer the reader to existing reviews for more details on the semion
superfluid.\refto{losalamos,braz,frad}.

Even with this limited glimpse at the semion superfluid, we see that the
Hall fluid, the chiral spin fluid, and the semion superfluid are intimately
related. Here I focus on the Hall fluid and treat the other two fluids as a
sort of afterthought. In \Ref{braz} I do the reverse, emphasizing the chiral
spin fluid and the semion superfluid and treating the Hall fluid almost as a
footnote.

We started this section by asking whether or not Mother Nature knows
about the even integers. Indeed, in \(4.5) and its subsequent generalization
we have found topological fluids described by
$$
\cL = 4({q\over 2}){1\over 8\pi} a\epsilon \part a
= {1\over 4\pi} q a\epsilon \part a
\eqno(even)
$$
with $q$ an even integer. The corresponding chiral spin fluid has statistics
angle
$\theta/\pi = 1/q =1/2, 1/4,$ ...{\bf .}  We find this an elegantly unified
picture
of the Hall fluid and the chiral spin fluid.  Indeed, were the chiral spin
fluid unknown it may have to be invented in order to fill in the ``gaps" at
the even integers.

Well, while the Hall fluid is realized experimentally, the chiral spin fluid
and the semion superfluid remain, alas,
at
this
point, figments of the theorists' imagination.  Nevertheless, the fact that the
physics and the mathematics of these  the structure are so intimately related
continues to sustain the optimists. One may be tempted  to
think that it would be ``unnatural'' for only one of them to be realized
experimentally.

\head {XV. Concluding Remarks and a Puzzle}
\taghead{15.}

It is now time for a few summarizing remarks. We have explored the
emerging subject of topological fluids.
One interesting issue is how to characterize and label the orders in these
systems. Traditionally, one uses broken symmetries and their associated
order
parametes to classify order and universality classes. Here, it appears that
these topological orders have to be characterized by the matrix $K$.

The physics we are getting at here originates from the very foundation of
quantum physics. As I emphasized in the introduction, the basic physics of
the Hall fluid originates from the interplay between fermi statistics and the
magnetic gauge force. Central to the physics of topological fluids is the
notion of the phase, perhaps the most startling and profound concept of
quantum physics, a concept totally and
categorically alien to classical physics. Already in the
Dirac-Aharonov-Bohm effect we see the emergence
of topology. When a charged particle
moves around a flux line, its wave function acquires a phase. This phase
does not depend on the precise shape of the closed curve traced out by the
particle, nor does it depend on how slowly the particle moves around the
curve. This is physics without rulers and without clocks.

Let me end this series of lectures with a story. A physicist was lucky
enough to have two girl friends, one named Lucy and one named Rita,
whom he liked equally. (I am afraid that this story is not going to be
politically correct.)
The physicist lived on a train line precisely half-way
between the homes of Lucy and Rita. (Thus, he found it convenient to think
of everything happening in a (1+1) dimensional spacetime and of Lucy
and Rita as the ``left" and ``right" excitation respectively.) Since he could
never decide whom to visit, he simply left it up to chance. It so happened
that on this line trains went left and go right at precisely the same
frequency, once every hour. Our physicist thus went to the train station
whenever the mood struck him and took the first train that came along. He
figured that in the long run he would end up visiting Lucy just as often as
Rita. In fact, after some months he found that he was visiting Rita nine
times more often than Lucy! Unfortunately for our physicist, both women
got angry at him, Lucy for his not visiting her often enough, and Rita for
his hanging around her too often. He ended up with no girl friend.

1) What is the moral of the story?
2) How did it happen that parity was spontaneouly broken?
3) What does this story have to do with this lecture (and Wiegmann's
lecture)?

\head{Acknowledgement}
I would like to thank H. Geyer for inviting me to an exceptionally well run
and stimulating school and to visit South Africa at a historic time just
before its first election. His warm hospitality made it possible for me to see
many different aspects of South~Africa.
I would like to thank my various collaborators, including D.~Arovas,
J.~Fr\"ohlich, J.R. Schrieffer,
 F. Wilczek,  Y.S. Wu, and especially X.G. Wen.
 I would also like to
thank those condensed matter physicists who have taken the effort over the
years to educate me on various topics. These include I. Affleck, P.
Anderson, G. Baskaran, P.M.A. Fisher,
 H. Fukuyama, K. Ishikawa,  S. Kivelson, M. Kohmoto,  R. Laughlin, M.
Rice,
D.  Scalapino, R. Scalettar, J.R. Schrieffer, P. Wiegmann, and
many others.

\references

\refis{Bogo} N.N. Bogoliubov, {\sl J. Phys.} (USSR) {\bf 11}  23 (1947).




\refis{ISKY} N. Imai, K. Ishikawa, T. Matsuyama, and I. Tanaka, \prb 42,
 10610, 1990, and references therein.

\refis{TRUG} C.A. Trugenberger, \pl B288, 121, 1992; \pr D45, 3807, 1992.

\refis{LM}  J. M. Leinaas and J. Myrheim, {\sl Nuovo Cimento} {\bf
37B}, 1, (1977).

\refis{FS} F. Wilczek, \prl 48, 1144, 1982; \prl 49, 957, 1982.



\refis{WWZ} X.G. Wen, F. Wilczek, and A. Zee, {\sl Phys. Rev.} {\bf
B39}, 11413 (1990).

\refis{KMT} M. Kohmoto, \pr B39, 1989, 11943; Y. Hatsugai and M.
Kohmoto, \pr B42, 8282, 1990; and references therein.

\refis{WZIX} X.G. Wen and A. Zee, {\sl Nucl. Phy.} {\bf B326}, 619 (1989).

\refis{WZDD1} X.G. Wen and A.  Zee, \prl 62, 1937, 1989.

\refis{WZDD2} X.G. Wen and A.  Zee, \pr B41, 240, 1990;
 \journal Int. J. Mod. Phys., B4, 437, 1990.















\refis{losalamos} A. Zee,  in {{\it High Temperature
Superconductivity}}, edited by K. Bedell, D. Coffey, D. Pines and J.R.
Schrieffer (Addison-Wesley 1990).

\refis{braz} A. Zee, in {{\it Particle Physics}}, ed. O.~\'Eboli  \etal,
{{\it Physics in (2+1)-Dimension}}, ed. Y.M.~Cho, {{\it Cosmology and
Elementary Particles}}, ed. D.~Altschuler, World Scientific.

\refis{GP} {{\it The Quantum Hall Effect}}, edited by  R.E. Prange
and  S.M. Girvin (Springer-Verlag, New York 1987).

\refis{WZZ} F. Wilczek and  A. Zee, \prl 51,  2250, 1983.

\refis{ASWZ}  D. Arovas, R. Schrieffer, F. Wilczek and A. Zee,\np B251,
117, 1985.


\refis{WEZ}  X.G. Wen and A. Zee, \prl  63, 461, 1989.










\refis{TOS4} X.G. Wen and A. Zee, \prb 41, 240, 1990.

\refis{TOS1} A. Fetter, C. Hanna, and R. Laughlin, \prb 39, 9679, 1989.

\refis{TOS2} Y. Hosotani and S. Chakravarty, \prb 42, 342, 1990; J.E.
Hettrick, Y. Hosotani, and B.-H. Lee, {\sl Ann. Phys. (N.Y.),} {\bf 209}, 151
(1991); J. Hetrick and Y. Hosotani, \prb 45, 2981, 1992.

\refis{TOS3}  Y.H. Chen, F. Wilczek, E. Witten, and B.I. Halperin,
\journal Int.  J. Mod. Phys., B3, 1001, 1989.




\refis{TOS5} H. Mori, \pl A146, 335, 1990.




\refis{AM} I. Affleck and J.B. Marston, \prb 37, 3774, 1988.

\refis{K} G. Kotliar, \prb 37, 3664, 1988.

\refis{IL} L.B. Ioffe and A.I. Larkin, \prb 39, 8988, 1989.



\refis{FOT} X.G. Wen and A. Zee, \journal J. de Physique, 50, 1623,
1989.

\refis{KI} K. Ishikawa and T. Matsuyama, \journal Z. Phys., C33, 41,
1986.; \np B280, 523, 1987.






\refis{WUZ} Y.S. Wu and A. Zee, \np B258, 157, 1985; \np B272, 322,
1986.








\refis{HR} D. Haldane and E.H. Reyazi, \prl 54, 327, 1985; \prb 31, 2529,
1985.

\refis{ETHE} X.G. Wen and A. Zee, \prb 44, 274, 1991.

\refis{ZHG} S.C. Zhang, T.H. Hansson, and S. Kivelson, \prl 62, 82, 1989.

\refis{FZ} J. Fr\"ohlich and A. Zee, \np B364, 517, 1991.

\refis{WTF} X.G. Wen, \pr B40, 7387, 1989; \journal Int. J. Mod. Phys.,
B2, 239, 1990.

\refis{TF} E. Witten, {\sl Comm. Math. Phys.} {\bf 117}, 353 (1988);
  {\bf 121}, 351 (1989).

\refis{cs} texas conference.

\refis{ab} south carlina aharonov bohm conf.

\refis{CSTX}
   A. Schwarz, \journal Lett. Math. Phys., 2, 201, 1978.; \cmp 67, 1,
1979;
W. Siegel, \np B156, 135, 1979.

\refis{CSTY} J.F. Sch\"onfeld, \np B185, 157,
1981; S. Deser, R. Jackiw, and S. Templeton, \prl 48, 975, 1982.

\refis{HG} Y.S. Wu and A. Zee, \pl B207, 39, 1988.

\refis{J1} J.K. Jain, {\sl Phys. Rev. Lett.} 199, (1989).

\refis{GW} M. Greiter and F. Wilczek, \journal Mod. Phys. Lett. B, 4,
1063, 1990.

\refis{J2} J.K. Jain, \pr B40, 8079, 1989.


\refis{vortex} A. Zee, \np B421, 111, 1994.


\refis{TW} R. Tao and Y.S. Wu, {\sl Phys. Rev.} {\bf B39}, 1087 (1984).

\refis{HP} B. Halperin, {\sl Phys. Rev.} {\bf B25}, 2185 (1982).


\refis{NR} N. Read, \prl 65, 1502, 1990.

\refis {moon} K. Moon, H. Mori, K. Yang, S.M. Girvin, A. H.
MacDonald, L. Zheng, D. Yoshioka, and S. C. Zhang, Indiana preprint
1994.

\refis{B} G.S. Boebinger \etal, \prl 64, 1793, 1990.

\refis{T} Y.W. Suen \etal, \prl 68, 1379,  1992.

\refis{E} J.P. Eisenstein \etal, \prl 68, 1383, 1992.

\refis{YMG}  D. Yoshioka, A.H. MacDonald and S.M. Girvin,
\pr B39, 1932, 1989.

\refis{MYG}  A.H. MacDonald, D. Yoshioka, and S.M. Girvin,
\pr B39, 8044, 1989.

\refis{RM} M. Rasolt and A.H. MacDonald, \pr 34, 5530, 1986.

\refis{FG} H.A. Fertig, \prb 40, 1087, 1989.

\refis{MPB} A.H. MacDonald, P.M. Platzman, and G.S. Boebinger,
\prl 65, 775, 1990.

\refis{R} L. Brey, \prl 65, 903, 1990.

\refis{Hlp} B.I. Halperin, \journal Phys. Helv. Acta., 56, 75, 1983.


\refis{WZ2} X.G. Wen and A. Zee, \pr B41, 240, 1990.

\refis{Wil} F. Wilczek, \prl 69, 132, 1992.



\refis{pkv} A. Polyakov, \np B120, 429, 1977.

\refis{WZR}   X.G. Wen and A. Zee, {\it Nucl. Phys.\/}, (suppl.) {\bf B15}, 135, (1990).

\refis{bw} B. Blok and X.G. Wen, \prb 42, 8133, 1990; \prb 43, 8337, 1991.

\refis{effective} X.G. Wen and A. Zee, \prb 44, 274, 1991.

\refis{ab} {{\it Quantum\ Coherence\/}}, edited by J. S. Anandan, World
Scientific 1990.

\refis{ezawa} Z.F.  Ezawa and A. Iwasaki, \prb 43, 2637, 1991.

\refis{FKT} J. Fr\"ohlich and T. Kerler, \np B354, 369, 1991.

\refis{cs} {{\it Physics\ and\ Mathematics\ of\ Anyons\/}}, edited by S. S.
Chern \etal, World Scientific 1990.

\refis{gir} S. Girvin and A.H. MacDonald, \prl 58, 1252, 1987, S. Girvin, in
{{\it The\ Quantum\ Hall\ Effect\/}}, edited by R.E. Prange and S.M.
Girvin (Springer-Verlag, New York 1987).

\refis{wniu} X.G. Wen and Q. Niu, \prb 41, 9377, 1990.

\refis{mpaf} M.P.A. Fisher and D.H. Lee, \prl 63, 903, 1989; D.H. Lee
and M.P.A.  Fisher,  \journal Int. J. Mod. Phys, B5, 2675, 1991.

\refis{wzmono1} X.G. Wen and A. Zee, \prl 69, 1811, 1992.

\refis{wzmono2} X.G. Wen and A. Zee, \prb 47, 2265, 1993.

\refis{wzmono3} X.G. Wen and A. Zee, ``A Phenomenological Study of
Interlayer Tunnelling in Double-Layered Quantum Hall Systems," MIT-
ITP preprint, 1994.


 \refis{yang} K. Yang, K. Moon, L. Zheng, A. H. MacDaonald, L. Zheng,
D. Yoshioka, and S. C. Zhang, \prl 72, 732, 1994.

\refis{murphy} S. Q. Murphy, J. P. Eisenstein, G. S. Boebinger, L.N.
Pfeiffer, and K. W. West, \prl 72, 728, 1994.

\refis{haldane} I. Tamm, \journal Z. Phys., 71, 141, 1931;
F.D.M. Haldane, \prl 51, 605, 1983.

\refis{wz2290} X. G. Wen and A. Zee, \prb 46, 2290, 1992.

\refis{shift} X. G. Wen and A. Zee, \prl 69, 953, 3600(E), 1992.

\refis{dhwen} D.H. Lee and X.G. Wen, \prb 49, 11066, 1994.

\refis{ctz} A. Cappelli, C. A. Trugenberger, and G. R. Zemba, \prl 72,
1902, 1994.

\refis{ich} I. Ichinose and T. Ohbayashi, Tokyo preprint (1994).

\refis{frad} E. Fradkin, {{\it Field Theories of Condensed Matter
Systems\/}},
Addison-Wesley Publishing Co. 1991.

\refis{haldaneh} F. D. M. Haldane, \prl 51, 605, 1983.

\refis{halperinh} B.I. Halperin, \prl 52, 1583, 2390(E), 1984.

\refis{laughlinh} R. B. Laughlin, \journal Surface Science, 141, 11, 1984.

\refis{ando} T. Ando, Y. Matsumoto, and Y. Uemura,  {\it J. Phys.
Soc. Japan} {\bf 39}, 279, 1975.

\endreferences

\end